**Beyond Seamless: Unexpected Defective Merging in Single-Orientation Graphene**

*Zhien Wang[1,2,†], Jiangtao Wang[2,*,†], Diego Exposito[3], Andrey Krayev[4], Shih-Ming He[2], Xudong Zheng[2], Zachariah Hennighausen[2], Ivan Brihuega[3,5], Se-Young Jeong[6,7,*], Jing Kong[2,*]*

[1]Department of Materials Science and Engineering, Massachusetts Institute of Technology, Cambridge, MA 02139, United States

[2]Department of Electrical Engineering and Computer Science, Massachusetts Institute of Technology, Cambridge, MA 02139, United States

[3]Departamento de Física de la Materia Condensada, Universidad Autónoma de Madrid, E-28049 Madrid, Spain

[4]HORIBA Scientific, HORIBA Instruments Incorporated, 359 Bel Marin Keys Blvd, Suite 18, Novato, CA, 94949, USA

[5]Condensed Matter Physics Center (IFIMAC), Universidad Autónoma de Madrid, E-28049 Madrid, Spain

[6]Department of Optics and Mechatronics, Pusan National University, Busan 46241, South Korea

[7]Department of Physics, Korea Advanced Institute of Science and Technology (KAIST), Daejeon 34141, Republic of Korea.

*Corresponding authors: Jiangtao Wang, Se-Young Jeong, Jing Kong

Email: wangjt@mit.edu, syjeong@pusan.ac.kr, jingkong@mit.edu

† Zhien Wang (author1) and Jiangtao Wang (author2) contributed equally to this work.

Keywords: single-crystal graphene; single-orientation stitching; defective merging; water-permeable defects

## Abstract

Single-orientation stitching of graphene has emerged as the predominant method for growth of large-area, high-quality graphene films. Particularly noteworthy is graphene grown on single-crystalline Cu(111)/sapphire substrates, which exhibits exceptionally planar oriented stitching due to the atomically smooth substrate, facilitating the formation of continuous, high-quality graphene monolayer. These single-orientation stitches have conventionally been regarded as seamless with negligible defect concentrations. In this report, we present experimental observations regarding graphene grown on single-crystalline Cu(111)/sapphire substrates. Among the graphene flakes with single-orientation, our findings reveal two major merging behaviors: one producing the expected seamless stitching, and another unexpectedly generating structural defects that create nanoscale pathways permitting water permeation. Notably, we identify a unique merging structure—overlapped junction, in which the edge of one graphene flake overlaps and lies atop the edge of another flake, rather than forming a continuous atomic stitch. This discovery challenges the conventional anticipation of single-orientation stitched graphene films as seamless single crystalline film, while offers unique perspective for graphene applications in molecular sieving, selective filtration membranes, and protective coatings.

## 1. Introduction

Single-crystalline graphene has been highly desirable for studies in fundamental physics, and its applications in high performance devices. The most effective method for synthesizing large-area single-crystalline graphene is the seamless stitching of unidirectional graphene flakes via chemical vapor deposition (CVD). The alignment of graphene flakes is enabled by using single-crystal Cu(111) as the epitaxial growth substrate.[1–5] It is anticipated that single-crystal Cu(111)/sapphire, which is fabricated through electron-beam or sputter deposition, is a superior growth substrate than Cu(111) foil. Cu(111)/sapphire facilitates the elimination of wrinkles to produce graphene with ultra-flat nature, which is believed to reduce defect density and improve the quality of single-crystal graphene.[6,7] Structural defects such as Stones-Wales defect, point defects and line defects, could be formed during CVD graphene growth. These defects deteriorate graphene's mechanical strength[8], conductivity[9] and thermal stability [10]. However, the properties of defects in graphene add new functionalities for graphene-based devices in water desalination, quantum electronics, and electrochemical applications [11–13]. The most common defect in polycrystalline graphene is grain boundary, which is formed when graphene flakes of two different orientations stitch together. On the other hand, the stitches of single-orientation graphene flakes on atomic-smooth Cu(111)/sapphire substrate are generally perceived as seamless without formation of grain boundary, which is crucial for producing large-area, high-quality single-crystal graphene film. However, here we report the first observation of water-permeable structural defects at the merging region of single-orientation graphene flakes grown on single-crystalline Cu(111)/sapphire substrate. These unexpected structures are detected under scanning electron microscopy (SEM) and the phase profile of atomic-force microscopy (AFM), appearing as a merging line of clear contrast at the stitching area of two graphene flakes. Scanning tunneling microscopy (STM) further reveals that some of these merging lines are overlapped junctions, where one graphene flake edge lies atop another rather than forming a continuous stitch. Water permeability of such defects are revealed through wet oxidation of the copper substrate underneath graphene when the graphene flakes on Cu(111)/sapphire is immersed in hot water. This finding challenges the prevailing anticipation that single-orientation stitched graphene films function as uniformly impermeable barriers[14], while on the other hand, it reveals the potential applications of the stitched graphene in other applications such as separation and selective filtration membranes.

## 2. Results

We first developed growth recipe of ambient pressure CVD for graphene growth on 700nm Cu(111)/sapphire. The single-crystal Cu(111) film is grown by atomic sputtering epitaxy on sapphire, which gives atomic-smooth surface upon deposition (**Figure 1a**).[15] The growth of graphene is under ambient pressure to minimize the evaporation of copper under high temperature, details of the recipe is summarized in Figure 1b. The SEM, AFM images and Raman mapping of the graphene grown on Cu(111)/sapphire are included in supporting information. The Raman mapping over 40μm×40μm area has D/G peak intensity ratio close to 0, while 2D/G peak intensity ratio around 2.6, which indicates high quality of the as-grown graphene (**Figure S1**). The optical microscope image of the graphene flakes on Cu(111)/sapphire is shown in Figure 1a, which shows single-orientation aligned graphene flakes. The sample is slightly oxidized in the air to better visualize the graphene flakes on Cu(111). However, when we examined the merging region of the aligned graphene flakes under SEM, we observed two different patterns of the merging region. One is perfectly smooth without any wrinkles, as shown in Figure1c, the other shows a vague line at the merging region, even though two flakes are of the same orientation, as shown in Figure 1d. Similar phenomenon is also observed using AFM. Figure 1e shows the merging of three graphene flakes. The merging region on the left shows a merging line of clear contrast under AFM phase profile, while the merging region on the right does not. Similarly in Figure 1f, two flakes of the same orientation stitched together vertex-to-vertex, resulting in 60° angle. A line at the merging region is clearly visible under the phase profile of AFM. Interestingly, these merging lines have very vague contrasts under the height profile. We have measured the height across the merging line in both cases, the step height is less than 0.5nm (Figure1g). This eliminates the possibilities of these merging lines being wrinkles, since wrinkles normally have a clear contrast in height profile with the step height around 0.8nm to over 1nm (**Figure S2**). We collected the topography (Figure 1h) and the corresponding contact potential difference (CPD) images (Figure 1i) for a junction area of two aligned graphene flakes with a merging line. The CPD image shows a clear increase of the CPD value (about 100mV) over this merging line, which indicates the merging of these two grains here is not seamless stitching (Figure 1i). These two types of merging behaviors (one with merging lines, the other without) are frequently observed on graphene samples from multiple different batches (**Figure S3,S4**), which suggests the possibility of different stitching mechanisms for single-orientation graphene flakes.

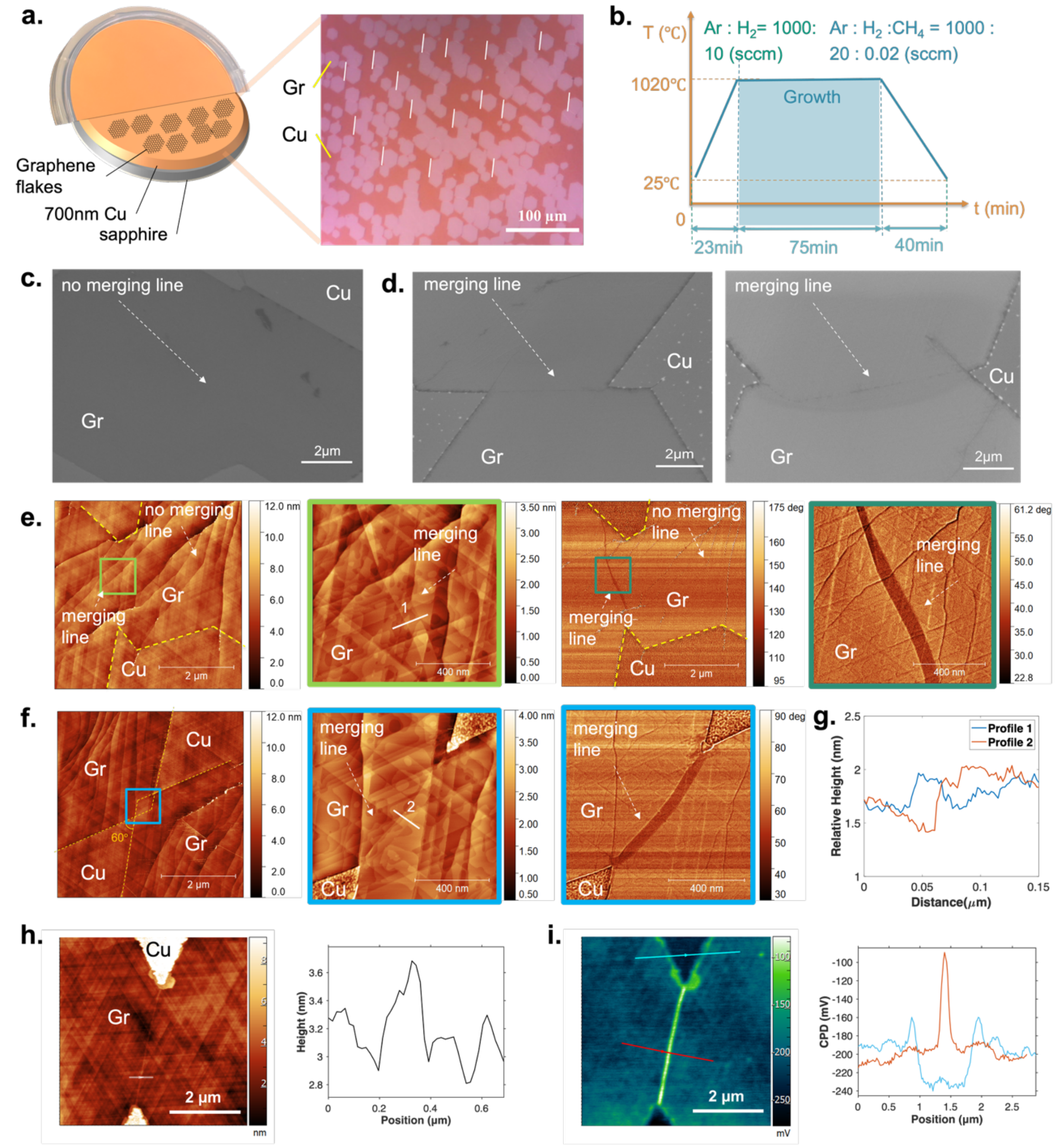

**Figure 1.** Characterizations of the merging line in single-crystal graphene. (a) Left: photograph of 2-inch 700nm single-crystal Cu(111) on sapphire with diagram illustrating the graphene-Cu(111)-sapphire structure; Right: Optical microscope image of graphene flakes with aligned orientation, growing on 700nm Cu(111)/sapphire (b) The growth recipe of graphene flakes on 700nm Cu(111) on sapphire (c) SEM of two merging graphene flakes without a merging line in between (d) SEM of two merging graphene flakes of aligned orientations with a merging line in between (e) AFM topography image of the merging region of graphene flakes, showing the merging line on the left side but not the right side (height profile and phase profile). Right: close-up of the height profile and phase profile of the merging line (f) AFM topography image of two merging graphene flakes of the same orientation, showing a merging line in between. The line has subtle contrast in the height profile but clearly visible in the phase profile. (g) AFM

height profile of the two merging regions. (h) Left: AFM topography image of a merging region of two aligned graphene flakes; Right: the height profile of the merging line. (i) Left: CPD image of the same area, indicating the merging of these two grains is not a seamless stitching; Right: the CPD section graphs along the correspondingly colored section lines in left panel

To investigate the property of the merging lines observed under SEM and AFM, we immersed the graphene flakes on Cu(111)/sapphire into 90℃ water for two hours right after growth. This is to visualize the diffusion of water through interfacial oxidation of Cu into $Cu_2O$.[16] If the merging line consists of defects, the Cu underneath it would also be oxidized and therefore have a color contrast after oxidation, since $Cu_2O$ is reddish in color under optical microscopy.[17] We have adjusted the temperature and time of oxidation such that graphene stays undamaged. The $Cu_2O$ underneath graphene can be detected using Raman, in which a sharp $Cu_2O$ peak is observed at around 644 cm-1 [16,18] , while graphene's G peak (1585 cm-1) and 2D peak (2705 cm-1) are both visible with 2D/G intensity ratio above 7, indicating that graphene is decoupled with water going underneath to oxidize the copper (**Figure 2a**, **Figure S5**) [19]. The Raman intensity map of $Cu_2O$ and 2D/G intensity ratio is obtained for three positions on the same sample. Raman maps of position and intensity of $Cu_2O$, G and 2D peaks (**Figure S6-S8**), and complete Raman spectrum of time-dependent oxidation of Cu(111) film (**Figure S9**) are included in supporting information. Position 1 shows the merging of three aligned graphene flakes. The merging region of the left two flakes has higher $Cu_2O$ intensity while the one on the right does not. But the 2D/G intensity ratio of both merging regions is roughly the same (Figure 2b middle and right panel). Similarly, position 2 shows the merging of a few aligned graphene flakes. The merging region on the left appears reddish under optical microscope imaging and shows higher $Cu_2O$ intensity. However, the one in the middle neither turns reddish nor has higher $Cu_2O$ intensity (Figure 2c). Position 3 shows the merging of roughly aligned graphene flakes. The two merging regions in the middle have Cu oxidized into $Cu_2O$, which is reflected by the color in the optical microscopic image and $Cu_2O$ Raman intensity map. While the merging region near the bottom does not contain $Cu_2O$ (Figure 2d). It is worth noticing that for $Cu_2O$ intensity maps in Figure 2, $Cu_2O$ has much higher intensity underneath graphene than the area without graphene coverage. This is explained by the graphene-enhanced oxidation process at the Cu-graphene interface, through the charge transfer between graphene and Cu [20].

Moreover, Figure2 b-d captured the speed of oxidation by correlating the amount of $Cu_2O$ formed with the relative position on graphene. It shows that water starts to diffuse from the edge of graphene, which causes the surrounding edges to be oxidized the most. In Figure 2c-d, water diffuses and oxidized to about 3μm from the edges for a duration of two hours. If there are no water-permeable defects, this should be the only way for water to go underneath graphene and oxidize the copper. However, Figure 2c-d present oxidation from grain boundaries and point defects, which indicate a second pathway: water-permeable defects. Based on the results, defective merging exists in the coalescence of aligned graphene flakes. To estimate the percentage of defective merging, we counted the number of defective merging lines versus seamless stitches in 145 merging flakes from the optical microscopy image of graphene flakes on Cu(111)/sapphire after water oxidation (Figure 2e, **Figure S10**). It shows that among 122 merging regions, 55 (around 45%) are defective merging lines while 67 (around 55%) merging regions are not oxidized, thus likely to be seamless stitching.

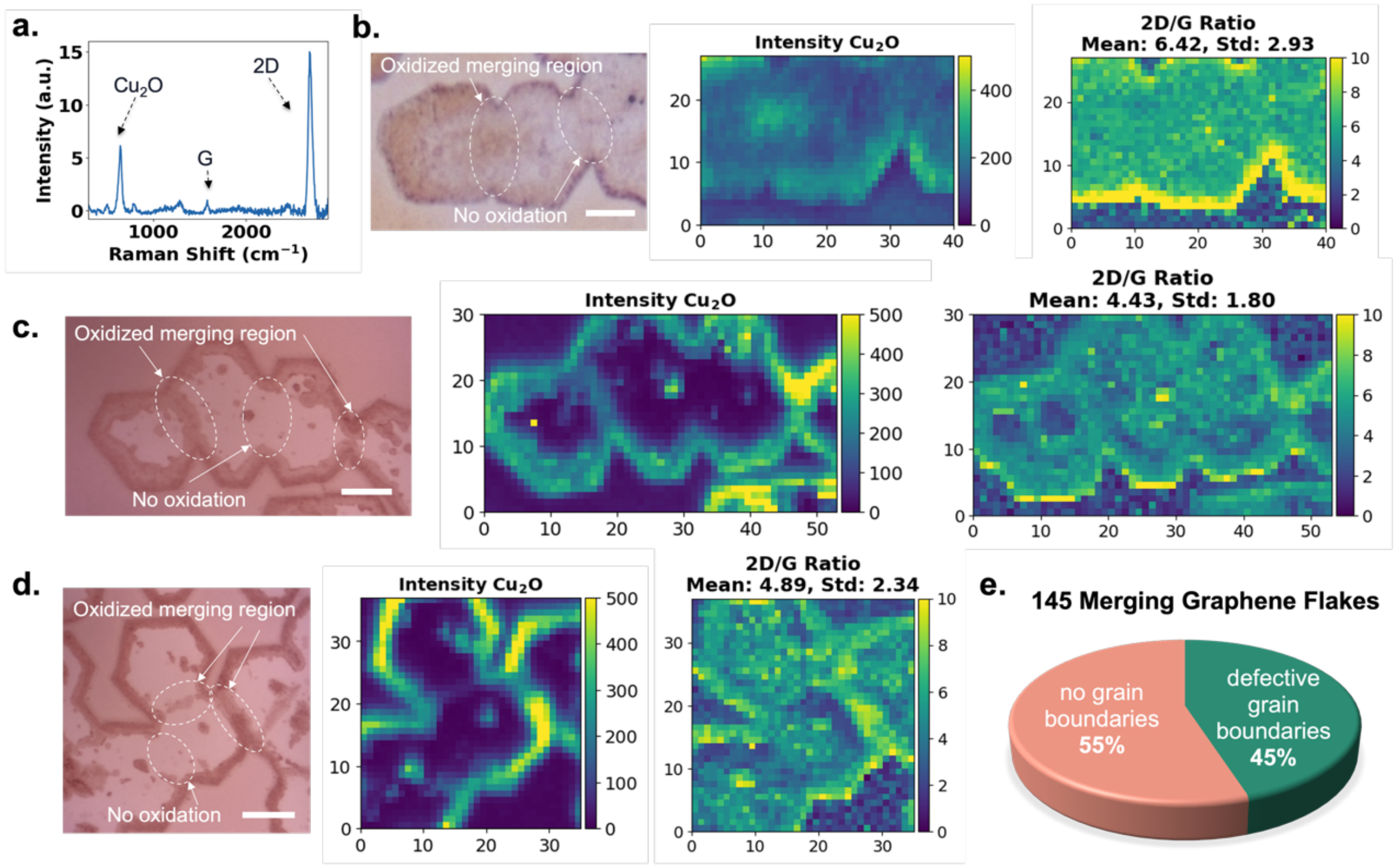

**Figure 2.** Raman mapping of the merging regions in single-crystal graphene on Cu(111)/sapphire after wet oxidation. (a) Raman spectrum of oxidized graphene flake on 700nm Cu(111) film/sapphire, showing a sharp $Cu_2O$ peak around 644cm$^{-1}$ using 473nm, 50mW laser. (b) Optical microscopic image of position 1 (more oxidized): merging graphene flakes after oxidation in 90°C water for two hours. Right: Raman mapping of $Cu_2O$ peak intensity and 2D/G intensity ratio over an area of 40μm×40μm. One of the merging regions shows higher $Cu_2O$ intensity while the other merging region does not. (c) Optical microscopic

image of position 2: merging graphene flakes with aligned orientation, after oxidation in 90℃ water for two hours. Right: Raman mapping of $Cu_2O$ peak intensity and 2D/G intensity ratio over an area of 52μm×52μm. One of the merging regions shows higher $Cu_2O$ intensity while the other merging region does not. (d) Optical microscopic image of position 3: merging graphene flakes with aligned orientation after oxidation in 90℃ water for two hours. Right: Raman mapping of $Cu_2O$ peak intensity and 2D/G intensity ratio over an area of 35μm×35μm. One of the merging regions shows higher $Cu_2O$ intensity while the other merging region does not. The degree of oxidation is slightly different for different regions on the copper film. (e) Statistics of the percentage of merging regions that show higher $Cu_2O$ signal vs. merging regions that do not show $Cu_2O$ signal in 145 graphene flakes. The scale bar in (b), (c) and (d) is 10μm, and the unit of the dimension marks in Raman map are all in μm.

To further investigate the orientation of graphene flakes on both sides of the stitches, we measured the atomic-resolution image on both sides of the graphene by conductive AFM at room temperature. The lattice orientation is compared by the extracted reciprocal lattice from Fast Fourier transform (FFT). **Figure 3a** shows an example of two graphene flakes with different orientations occurring on Cu(111) film, and a merging line is clearly visible in the phase profile while having little contrast in the height profile. However, in contrast, Figure 3b-c show two merging regions between graphene flakes, the FFT of graphene lattice from both sides shows aligned orientation. But similar merging lines in the phase profile of AFM in these two regions are also clearly seen. It is anticipated that when graphene flakes of different orientations are merged, the stitching could lead to defective boundaries, but the results here show that even for graphene flakes of the same orientation, merging between them could result in similar stitching boundaries (as revealed by the AFM height and phase profile), which is consistent with our observations shown in Figure 1 and 2.

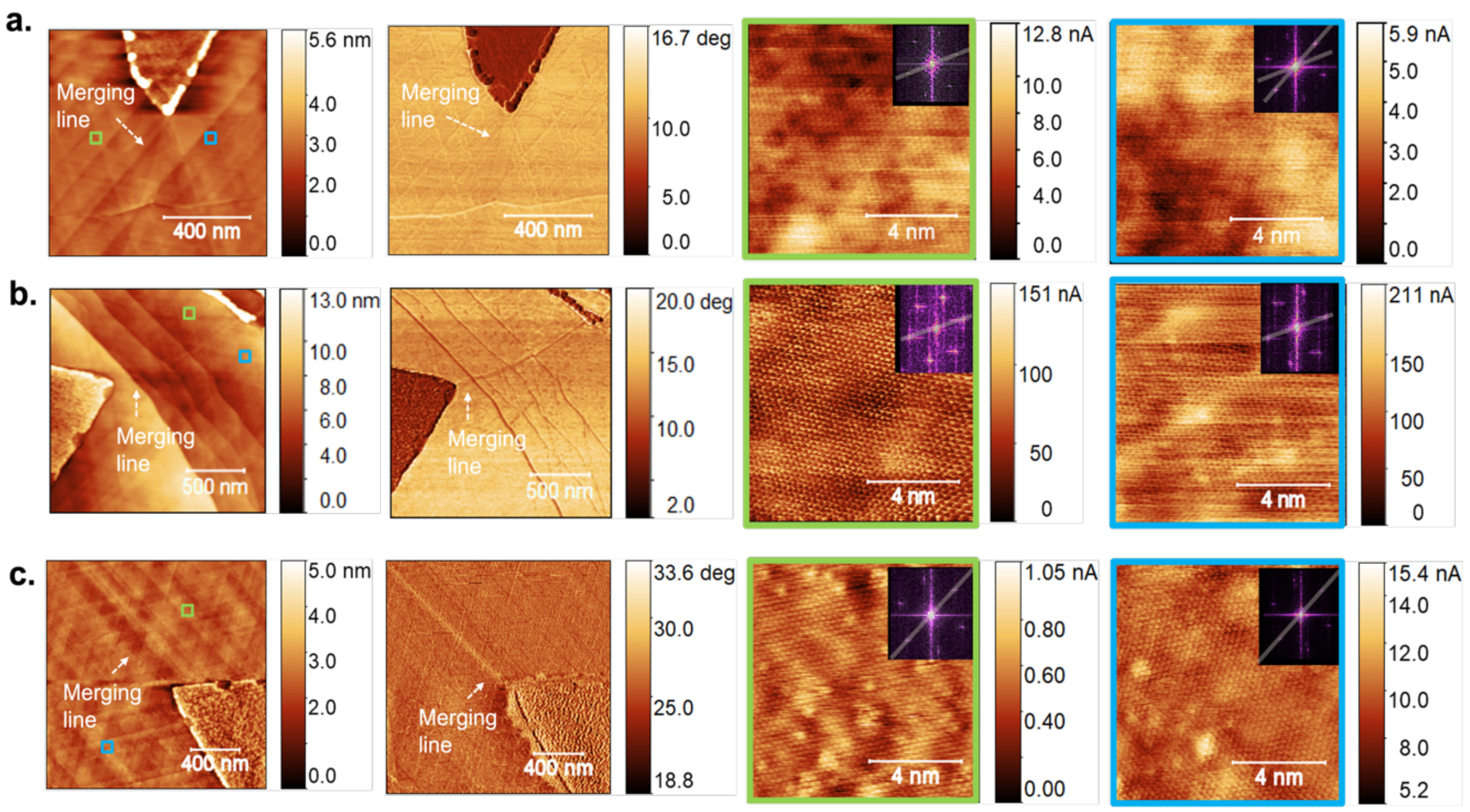

**Figure 3.** AFM of three merging regions and atomic resolution images of graphene lattice on two merging flakes, with FFT confirming the orientations. (a) AFM topography and phase profile at the merging region of two graphene flakes with different orientation. Right: atomic resolution cAFM of the two merging graphene flakes. The insets are FFT from the atomic-scale images. A line at the merging region with clear contrast is found from AFM phase profile. (b) (c) AFM topography and phase profile at the merging region of two graphene flakes with aligned orientation. Similar line structure is found at the merging region of two aligned graphene flakes. The green and blue square indicating the positions where atomic resolution images are taken. Right: atomic resolution conductive AFM (cAFM) of the two merging graphene flakes. The insets are FFT from the atomic-scale images. (b)and (c) both show aligned orientation of two merging flakes from the FFT.

To obtain a better insight into the structure of the merging line, we performed STM measurements over representative regions containing these boundaries, both for aligned and unaligned merging graphene flakes. We first identified specific merging regions having a merging line by optical microscopy and AFM at ambient conditions. Then, we characterized exactly the same graphene merging regions by STM at 4K in ultrahigh vacuum (UHV) conditions. STM images of the merging regions reveal the existence of a large variety of monoatomic steps across the merging lines. We identified their nature by correlating the measured step height in STM profiles with known values corresponding to the different step configurations (Gr-Gr step, Cu-Cu step, Gr-Cu step and combinations) (**Figure 4a**). We note that measured step heights should be interpreted with caution: STM measurements reflect a convolution of the actual topography and the local density of states (LDOS). The true topographic height is only reflected when the surface on both sides of the step is identical. If the surface nature differs, the measured step height may vary slightly depending on the tunneling conditions such as bias voltage, tunneling current, and the state of the tip apex.

Our STM measurements of merging lines between graphene flakes with aligned orientation (Figure 4), reveal the merging line as an overlapped graphene junction. Figure 4b-d presents the STM characterization of a merging line between two flakes with aligned orientation. The position of the merging line and the relative orientation of the flakes on both sides are confirmed by optical microscopy images and atomic-resolution STM images, detailed in **Figure S11**. If we follow the STM profile across the merging region (Figure 4c) from left to right we see that from A to B, we have a step height of +3.5Å, which corresponds to a height increase of one graphene layer (Gr-Gr step). In the STM image (Figure 4b), this edge of the merging line appears decorated by the presence of many brighter spots, which we ascribe to more reactive zones, such as dangling bonds from undercoordinated C atoms, at the termination of the top graphene layer. From B to C, we have a down step of -1.3Å, which corresponds to the height difference between a decrease of one graphene monolayer step (Gr-Gr step) and an increase of one underlaying copper step (Cu-Cu step).

Figure 4e-k shows STM measurements on another merging line between two single-orientation graphene flakes. Its position is shown in **Figure S12**, with Figure 4e corresponding to a magnified view of the merging line region outlined by an orange square in Figure S12e. Here again, the height profile across the merging line supports the presence of an overlapping region with one graphene layer on top of the other (Figure 4f-g). From A to B, we have a step height of +1.4Å, which corresponds to the height difference between an up step of one graphene layer (Gr-Gr step) and a down step of copper (Cu-Cu step). From B to C, we have a down step of -3.4Å, corresponding to a decrease of one graphene layer (Gr-Gr step). This edge of the merging line marks the termination of the upper graphene layer, appearing as brighter in the STM image (Fig 4e). From C to D, and D to E, we have up steps of +2.2Å, matching with underlying single copper steps (Cu-Cu step).

The overlapping nature of the merging line is further confirmed by the STM manipulation experiments shown in Figure 4h-k, performed on a distinct region of the same merging line, with its position marked by a dark blue square in Figure S12b. By scanning with the STM tip over a defective part of the edge of the merging line, we tore and fold the top graphene layer, locally exfoliating it (Figure 4h-i). This is similar to previous works, where they first created a defective graphene border and then tore and fold part of the graphene layer on itself [21,22]. Measuring the height of the folded graphene layer lying on top of the G/Cu(111) surface (blue line outlined in Fig 4i), we obtain a step height of ~3.3Å, which corresponds to the distance between two graphene layers (Figure 4j). Measuring the step height between the Cu surface region exposed after tearing graphene and the border of the merging line (green line outlined in Fig 4i), we get an up step of ~1.2Å (Figure 4k). This corresponds to the height difference between an underlaying Cu down step (Cu-Cu step) and a graphene up step (Gr-Cu step). The results align with the understanding of the existence of an overlapping area where the two graphene flakes meet, with one of the flakes on top of the other. This is consistent with the observed possibility of tearing the upper graphene layer from the defective border.

We also analyzed merging lines between two misaligned graphene flakes, such as the structure shown in Figure 3a. Our results show that these merging lines have a similar structure to the one observed in aligned flakes, with the presence of an overlapped graphene junction. In **Figure S13** we show a different merging region between two graphene flakes with a ~10° misorientation, estimated based on the edges of the graphene flakes (Figure S13a-c). This merging region also forms an overlapped junction, and tearing at the defective boundary occurred during scanning (Figure S13d–f), similar to what was observed in aligned flakes (Figure 4h-k). The crystallographic orientations of the flakes on both sides are confirmed by the atomic-resolution imaging and FFT analysis (Figure S13g–h).

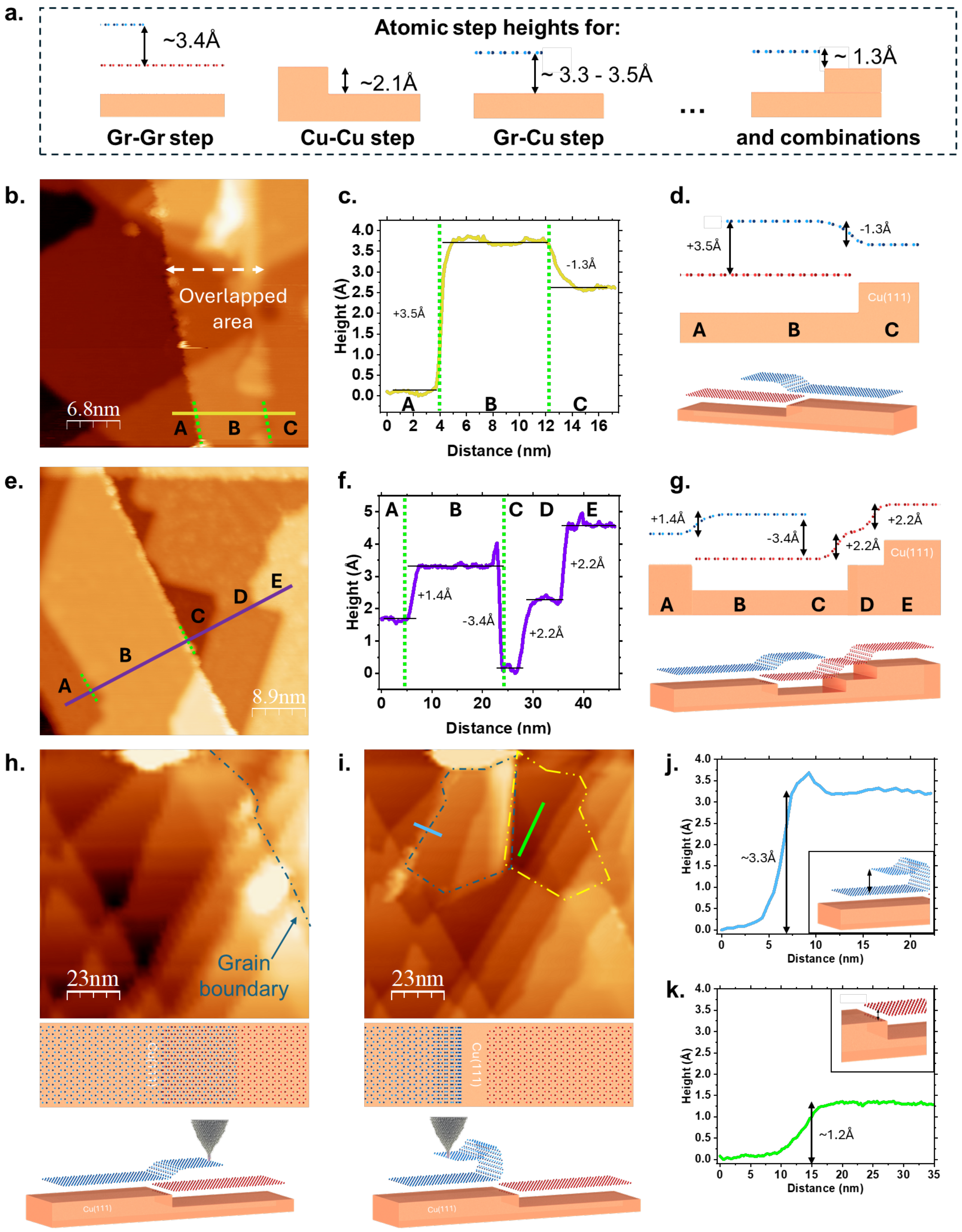

**Figure 4.** STM images of three merging regions, revealing the overlapped junction at the merging front. (a) Illustration of atomic step heights for Gr-Gr, Cu-Cu, Gr-Cu and combinations of Gr-Cu steps with Cu-Cu steps underneath. (b) STM image of one overlapped region between aligned merging graphene flakes. (c) Height profile across the overlapped region in (b). (d) Side view illustration corresponding to its morphology and 3D illustration. (e) STM image of a different overlapped region between aligned merging graphene flakes. (f) Height profile across the overlapped region in (e). (g) Side view illustration corresponding to its morphology and an angled illustration. (h) STM image of a different zone of the same merging area in (e), with top and side illustrations. The defective border of the top graphene layer on the overlapped region is marked with a blue dashed line. (i) Tearing of the graphene flake at a particularly defective zone of the boundary by simply scanning the region, with top and side illustrations. (j) Height profile across the folded graphene flake with the step height of Gr-Gr, inset shows the structure. (k) Height profile from the Cu surface region exposed after tearing graphene and the right Gr-Cu layer of the overlapping region, inset shows the structure.

## 3. Discussion

Through experimental characterizations and correlation with established literature, we propose a classification scheme for the merging behavior of graphene flakes during CVD growth on Cu(111) substrates. As summarized in **Table 1**, the merging behaviors can be grouped as the following: Type A: Seamless Stitching — Two graphene flakes of identical orientation merge coherently, forming a continuous lattice without introducing structural defects. Type B: Defective Stitching — Structural defects such as Stone-Wales (SW) defects, point defects etc. are introduced at the stitching of two misaligned graphene flakes. Type C: Overlapped Junction — A vertical overlap is formed where one flake terminates at a Cu step and the other overgrows it, resulting in a bilayer junction rather than in-plane stitching (see Figure 4a–k, Figure S11-S13).

| Label | Structure | Orientation | Type | Percentage |
|---|---|---|---|---|
| A | 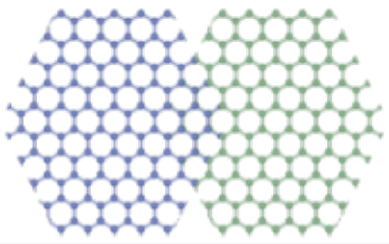 | Same | Perfect stitch | 55% |
| B | 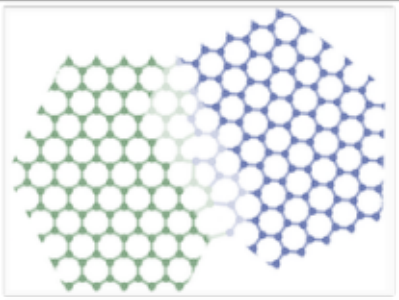 | Different | Defective Stitch | 45% |
| C | 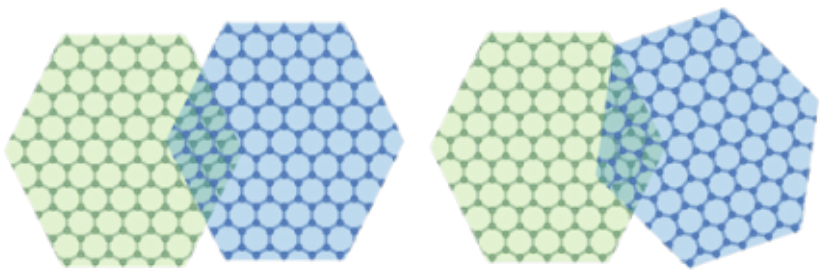 | Same or Different | Overlapped junction | |

**Table 1.** Classification of stitching in graphene flakes. A is the perfect stitching of two graphene flakes with aligned orientation, without any defects in the stitching area. This takes up around 55% from the calculation of not-oxidized merging region in wet oxidation experiment. B and C are the merging region with defects. This takes up around 45% from the calculation of oxidized merging region in wet oxidation experiment. B is the defective stitching, in which Stone-Wales (SW) defect, single or double vacancies or other types of defects could be present in the merging region, forming the grain boundary. While C is the formation of overlapped junction, which exists in the merging of graphene flakes with same and different orientation.

While the formation of grain boundaries in polycrystalline or misoriented graphene flakes is well studied, the presence of structural defects at the merging region of aligned graphene flakes on single-crystal Cu(111) remains less understood. In our study, firstly, we find that not all graphene flakes grown on Cu(111) share perfectly identical orientations—misalignments can occur despite the use of a single-crystalline substrate. While Cu(111) offers a well-defined crystallographic template, possibly the weak graphene–copper interaction, local surface inhomogeneities, and stochastic nucleation could lead to orientation mismatches between adjacent graphene flakes. As a result, the merging region can host various types of structural defects. One possibility is the Stone-Wales (SW) defects, which consists of five-seven member rings, and have been observed in single-crystal graphene [23]. One other possibility is the point defect, in which one or two atoms are missing at the merging region. A line of point defects have been observed in the single-orientation WS2 film,[24] similar structure is highly possible in stitched graphene grown on Cu(111).

Beyond the defect types expected from misorientations, we observed the overlapped junction (Type C in Table 1) at the merging region of both aligned and misaligned graphene flakes. Similar structure has been reported only in polycrystalline graphene grown on copper foil, as grain boundaries in misaligned graphene domains [25–29]. Misorientations of 10 to 25 degrees and high hydrogen partial pressure during growth have been identified as conditions where overlapped junctions are tend to form [27]. However, here we found the overlapped junction in merging of two graphene flakes without any measurable misalignment (Figure 4b, Figure S11). By directly characterizing this structure on its grown substrate, our results suggest the role of a copper step (~2.2 Å) in the formation of the overlapped junction. When a Cu step edge is present at the merging front between two graphene flakes, instead of forming an atomic stitch, one graphene flake terminates at the step while the adjacent flake continues to grow over it, resulting in a vertical overlap region approximately 10–50 nm wide. In both merging regions of aligned and misaligned graphene flakes, the overlapped regions are confirmed by STM height profiles and atomic-resolution imaging (Figure 4a–i, Figure S11-S13). These structures are characterized by asymmetric step heights (~1.2–1.4 Å on one side, ~3.3–3.5 Å on the other), consistent with a monolayer graphene flake stacked on another layer atop a Cu step. With the explanation of the formation of overlapped junctions, one might assume that these structures only occur when the Cu step edge is aligned with the direction of flake merging. However, we have observed merging lines forming at various angles relative to the direction of bunched Cu steps. The merging lines are perpendicular to the bunched Cu steps in Figure 3a, c, while nearly parallel in Figure 3b. In Figure 1e-f, the merging lines form in an angle with the bunched Cu steps. More examples are included in the supporting information (**Figure S14**). Although we do not have detailed STM data for these specific regions, some merging lines exhibit width and height contrasts in AFM images consistent with overlapped junctions. Since the Cu step under overlapped junction is only ~2.2 Å, it is possible that the detailed step structure of the Cu surface could be more complex than the bunched steps of almost a nanometer high. Based on our estimation from the wet-oxidation experiment (Figure2), approximately 55% of merging regions correspond to seamless stitches, while the remaining 45% are likely to be defective grain boundaries and overlapped junctions. Both defective grain boundaries and overlapped junctions are structurally distinct from seamless stitching and can introduce pathways for molecular transport. These structures can provide permeability to small molecules and ions. SW defects are anticipated to have extraordinary proton permeability and isotope selectivity.[30] It has been demonstrated that CVD graphene has high proton conductivity and is highly selective towards H+/D+ isotope while maintain negligible transport of bigger ions such as K+ (Potassium ion) [31]. Besides intrinsic defects such as SW defects, nanoscale wrinkles and ripples cause accumulation in strain, which enhance proton transport even in otherwise defect-free graphene [32]. We have also observed higher compressive strain at the merging region of single-orientation graphene flakes (**Figure S15**), which can be attributed to formation of defects or ripples, or both since structural defects cause buckling of graphene.[33] Regarding water permeability observed in oxidation of copper as shown in Figure2, the van der Waals diameter of a water molecule is around 0.275nm, which is the minimum size of water-permeable pore [34]. It is reported that defects and ripples in graphene facilitate the interfacial oxidation of copper [16]. These insights help to explain the water-permeable behavior observed at merging lines in wet-oxidation experiment and reinforce the role of structural imperfections—whether in the form of defective stitches or overlapped junctions—in governing functional properties of single-crystal graphene films.

## 4. Conclusion

Challenging the conventional understanding of single-orientation stitched graphene as a continuous, impermeable barrier, we reported the observation of nanoscale permeable structural defects at the merging region between aligned graphene flakes grown on single-crystal Cu(111)/sapphire. Through combined AFM, SEM, Raman, and STM characterizations, we identify and classify two major merging behaviors at the coalescence of aligned graphene flakes: (1) seamless stitching with negligible defect concentration; (2) formation of structural defects, including overlapped junctions where one graphene flake edge lies atop another during merging. Seamless stitching is distinguishable based on SEM images and AFM phase profiles, in which no merging line could be observed between two coalescing flakes, in contrast with the merging line observed at other coalescing region. STM further revealed the structure of some merging lines to be an overlapped junction, in which two graphene edges stacks atop of one another rather than stitches to be a continuous film. The wet-oxidation process also reveals the defective structures through interfacial oxidation of copper into $Cu_2O$. Nearly half of the merging region show high intensity of $Cu_2O$ in Raman intensity map, which indicates the presence of nanoscale water-permeable pathways at these regions. This finding advances the understanding of the single-orientation stitched graphene. On one hand, it is not defect-free as former anticipation assumed, on the other hand, these unexpected defective structures may find important use in nanofiltration, water desalination, and molecular sieving, etc., for industrial and environmental applications.

## 5. Experimental Methods

*CVD Graphene Growth*: Graphene was synthesized via ambient-pressure chemical vapor deposition (APCVD) on 700nm single-crystal Cu(111) films epitaxially grown on α-$Al_2O_3$(0001) sapphire substrates. The growth was carried out in a 1-inch quartz tube furnace. The single-crystal Cu(111)/sapphire substrate is fabricated by atomic sputtering, resulting in atomic-smooth single-crystal copper [15]. The substrates were loaded into the center of the furnace and heated to 1020 °C under a flow of 100 sccm Ar and 10 sccm $H_2$. Growth was initiated by introducing Ar: H2: $CH_4$ =1000:20:0.02 (sccm). The growth duration ranged from 75 to 120 minutes depending on the desired graphene coverage. After growth, the furnace was rapidly cooled to room temperature under Ar and $H_2$ flow.

*Raman Spectroscopy and Fitting*: Raman measurements were conducted using a Renishaw InVia Reflex Raman Confocal Microscope with a 532 nm, 50 mW laser, a 1200 l/mm holographic grating, and a 50× objective, yielding a beam spot size of ~1 μm. The incident laser power was set to 20 mW. Peak fitting was performed using a custom Python-based analysis script. Spectral background subtraction was carried out with the BaselineRemoval package [35]. Raman peaks were modeled using a Voigt profile function from the SciPy library, which incorporates four fitting parameters: peak center (Raman shift), standard deviation of the Gaussian component, half-width at half-maximum (HWHM) of the Lorentzian component, and peak intensity. Parameter optimization was achieved using the least-squares fitting algorithm also provided by SciPy.

*Conductive AFM Imaging*: Atomic force microscopy (AFM) measurements were performed using a Cypher VRS system. Atomic-resolution imaging was achieved using an ElectriMulti75-G probe, with a bias voltage of 0.5–5 mV applied between the tip and the sample. Scans were conducted over a 10 nm × 10 nm area at a resolution of 512 × 512 pixels and a scanning rate of 80 Hz. The FFT analysis was performed based on the atomic-resolution image to determine the orientation of graphene flakes.

*CPD Imaging*: The CPD images were collected on SmartSPM (HORIBA Scientific) using the frequency modulated Kelvin probe microscopy imaging mode with Access-SNC-GG probes (APPNano).

*STM Imaging*: The sample was annealed in UHV, first at 370K for 30min, and then at 470K for 25min more. Then, it was transferred to the STM keeping it all the time in the same ultra-high-vacuum (UHV) environment. STM measurements were performed with a homemade STM in UHV at a temperature of 4K. The STM microscope is designed to optimize the optical access to transferred samples and the capability of positioning the STM tip on the desired sample region within 5μm error. This enables the efficient mapping and identification of specific sample regions previously characterized by different external techniques such as AFM or optical microscopy. STM data was acquired and processed using the WSxM software [36].

**Acknowledgements**

This work was supported in part by the Semiconductor Research Corporation (SRC) Center 7 in JUMP 2.0 (Award No. 145105-21913). J.W. and J.K. acknowledge funding from the Air Force Office of Scientific Research (AFOSR) under the Multi-University Research Initiative FA9550-22-1-0166. D.E. and I.B. acknowledge financial support from the Spanish Ministry of Science and Innovation, through project PID2023-149106NB-I00, and María de Maeztu Program (CEX2023-001316-M), the Comunidad de Madrid and the Spanish State through the Recovery, Transformation and Resilience Plan "Materiales Disruptivos Bidimensionales (2D)" MAD2D-CM-UAM. X. Z. and J. K. acknowledge the support by the US Army Research Office grant number W911NF2210023. Z.H. and J. K. acknowledge the support from the U.S. Army DEVCOM ARL Army Research Office through the MIT Institute for Soldier Nanotechnologies under Cooperative Agreement number W911NF-23-2-0121 and S.-Y. J. acknowledges the support by the Basic Science Research Program through the National Research Foundation of Korea (NRF) funded by the Ministry of Science, ICT & Future Planning (RS-2024-00455226). This work utilized the Shared Experimental Facilities, partially supported by the National Science Foundation through the MRSEC Program (Award No. DMR-1419807), as well as the MIT.nano facilities. The authors acknowledge helpful insights from Prof. Feng Ding regarding this work.

# Supporting Information

## Beyond Seamless: Unexpected Defective Merging in Single-Orientation Graphene

*Zhien Wang[1,2,†], Jiangtao Wang[2,*,†], Diego Exposito[3], Andrey Krayev[4], Shih-Ming He[2], Xudong Zheng[2], Zachariah Hennighausen[2], Ivan Brihuega[3,5], Se-Young Jeong[6,7,*], Jing Kong[2,*]*

[1]Department of Materials Science and Engineering, Massachusetts Institute of Technology, Cambridge, MA 02139, United States

[2]Department of Electrical Engineering and Computer Science, Massachusetts Institute of Technology, Cambridge, MA 02139, United States

[3]Departamento de Física de la Materia Condensada, Universidad Autónoma de Madrid, E-28049 Madrid, Spain

[4]HORIBA Scientific, HORIBA Instruments Incorporated, 359 Bel Marin Keys Blvd, Suite 18, Novato, CA, 94949, USA

[5]Condensed Matter Physics Center (IFIMAC), Universidad Autónoma de Madrid, E-28049 Madrid, Spain

[6]Department of Optics and Mechatronics, Pusan National University, Busan 46241, South Korea

[7]Department of Physics, Korea Advanced Institute of Science and Technology (KAIST), Daejeon 34141, Republic of Korea.

*Corresponding authors: Jiangtao Wang, Se-Young Jeong, Jing Kong

Email: wangjt@mit.edu, syjeong@pusan.ac.kr, jingkong@mit.edu

**Graphene Growth and Characterizations.** We developed the recipe for ambient pressure (AP) CVD of graphene on Cu (111) film (Figure S1a). The evaporation of copper film is much less in ambient pressure and the graphene grain size can be up to 100μm. The AFM images (Figure S1b), SEM image (Figure S1c), and atomic resolution cAFM image (Figure S1d) were acquired directly from graphene on Cu (111) film. The atomic resolution cAFM image shows a clear atomic structure of graphene. Raman mapping is obtained on transferred graphene. FigureS1 e-f show 2D/G intensity ratio around 2.6 and D/G intensity ratio around 0. The Raman scattered plot shows that the transferred graphene is nearly doping-free (Figure S1f). All the characterization results indicate high quality of the graphene grown on Cu(111) film by ambient pressure CVD.

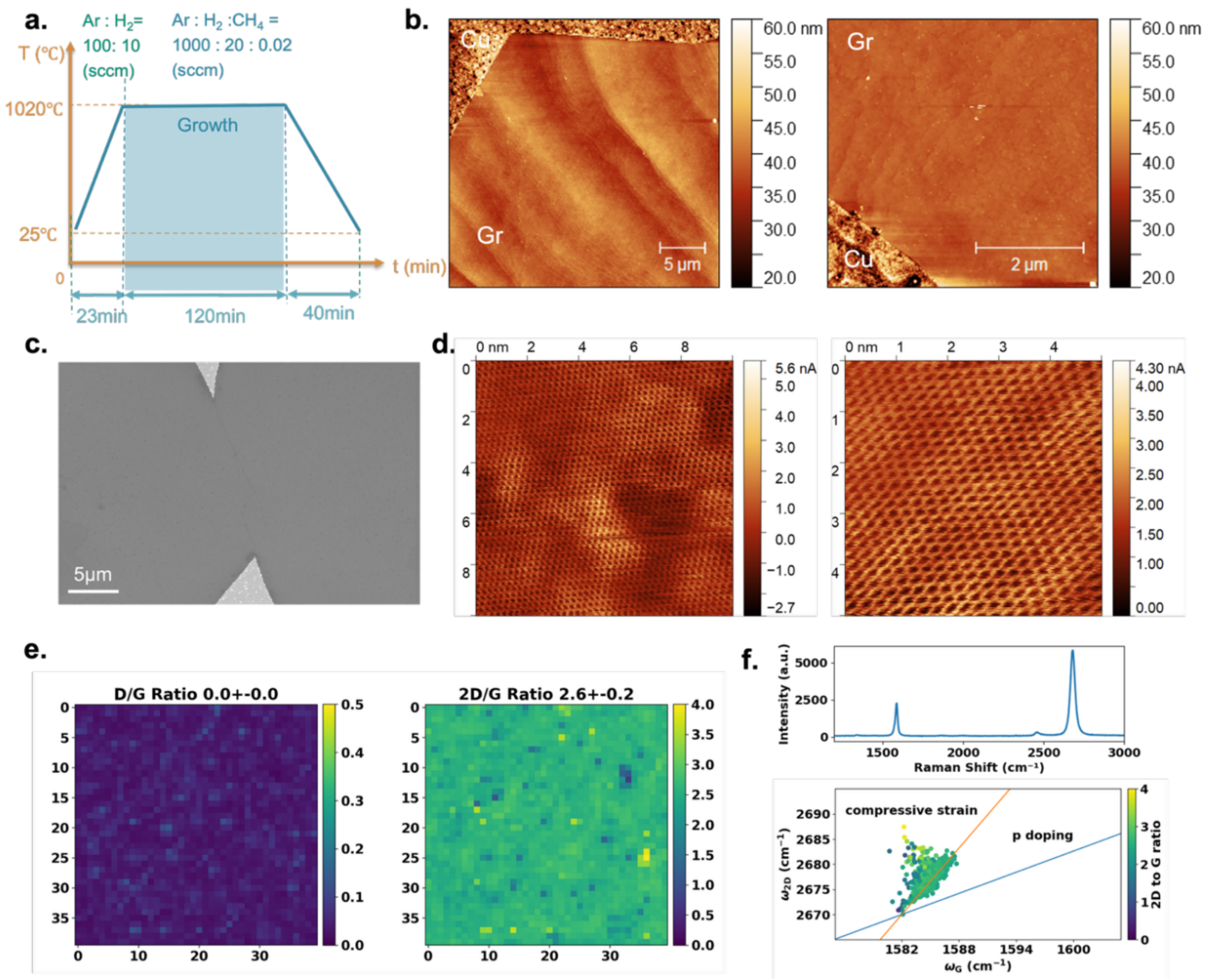

**Figure S1.** (a) CVD recipe of AP graphene growth on 700nm Cu(111) film (b) AFM images of as-grown graphene on Cu(111) film. Cu(111) film is oxidized after growth to locate the graphene flake under optical microscopy (c) SEM image of as-grown graphene on Cu(111) film (d) Atomic resolution cAFM image of graphene. (e) Raman mapping of D/G intensity ratio and 2D/G ratio for 40µm×40µm graphene transferred on SiO2 (f) Top: a typical Raman spectrum obtained from the sample; Bottom: Raman scattering plot of 1600 points, plotted along with the compressive strain and p doping line.

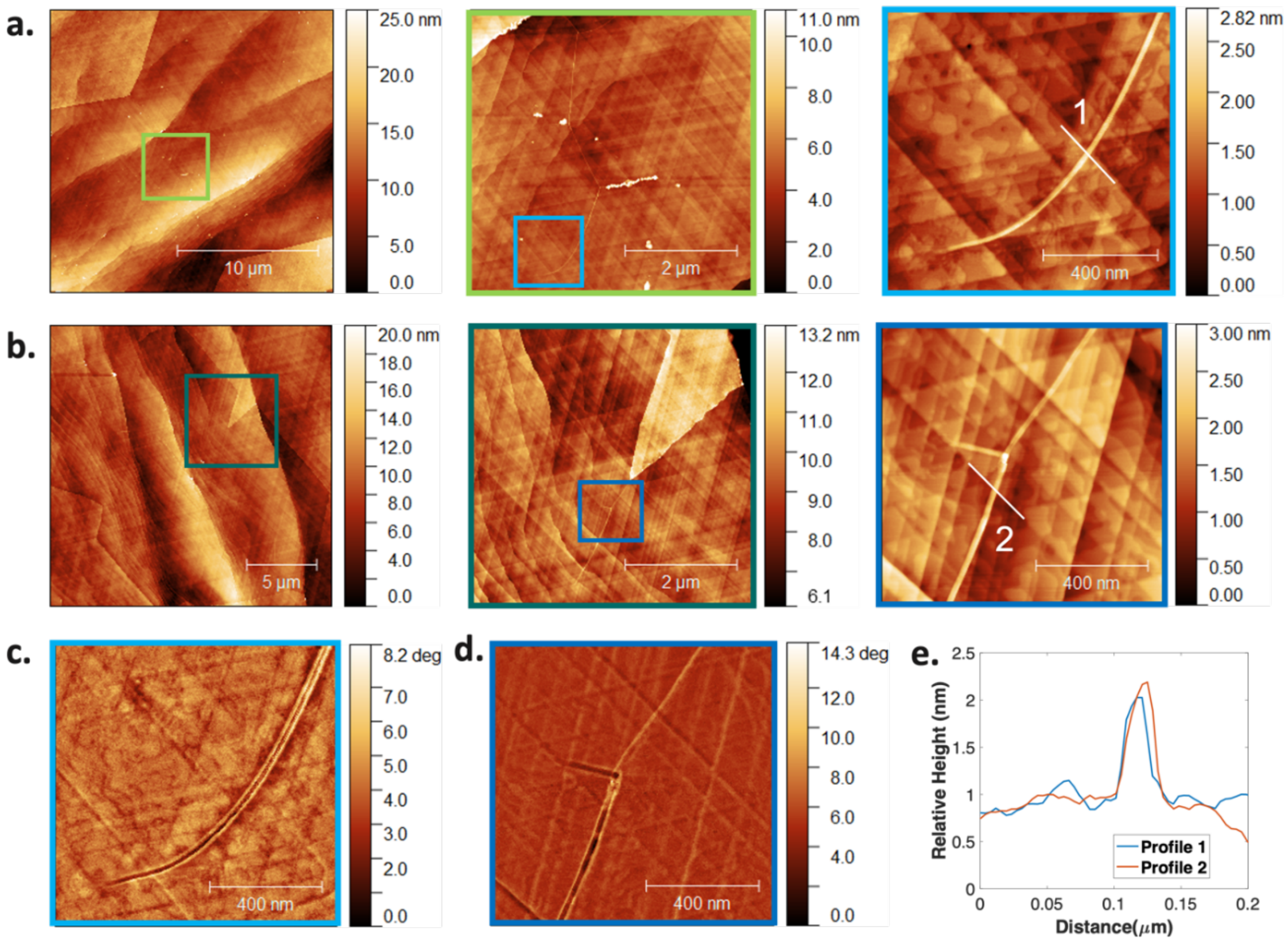

**Figure S2.** AFM profiles of wrinkles on graphene. (a) AFM topography profile of a wrinkle at the merging region of two flakes (indicated by the light green box). (b) AFM topography profile of a wrinkle on another merging region (indicated by the dark green box). (c) AFM phase profile of the wrinkle in (a). (d) AFM phase profile of the wrinkle in (b). (e) Height measurement of wrinkles in (a) and (b). Both are > 1nm.

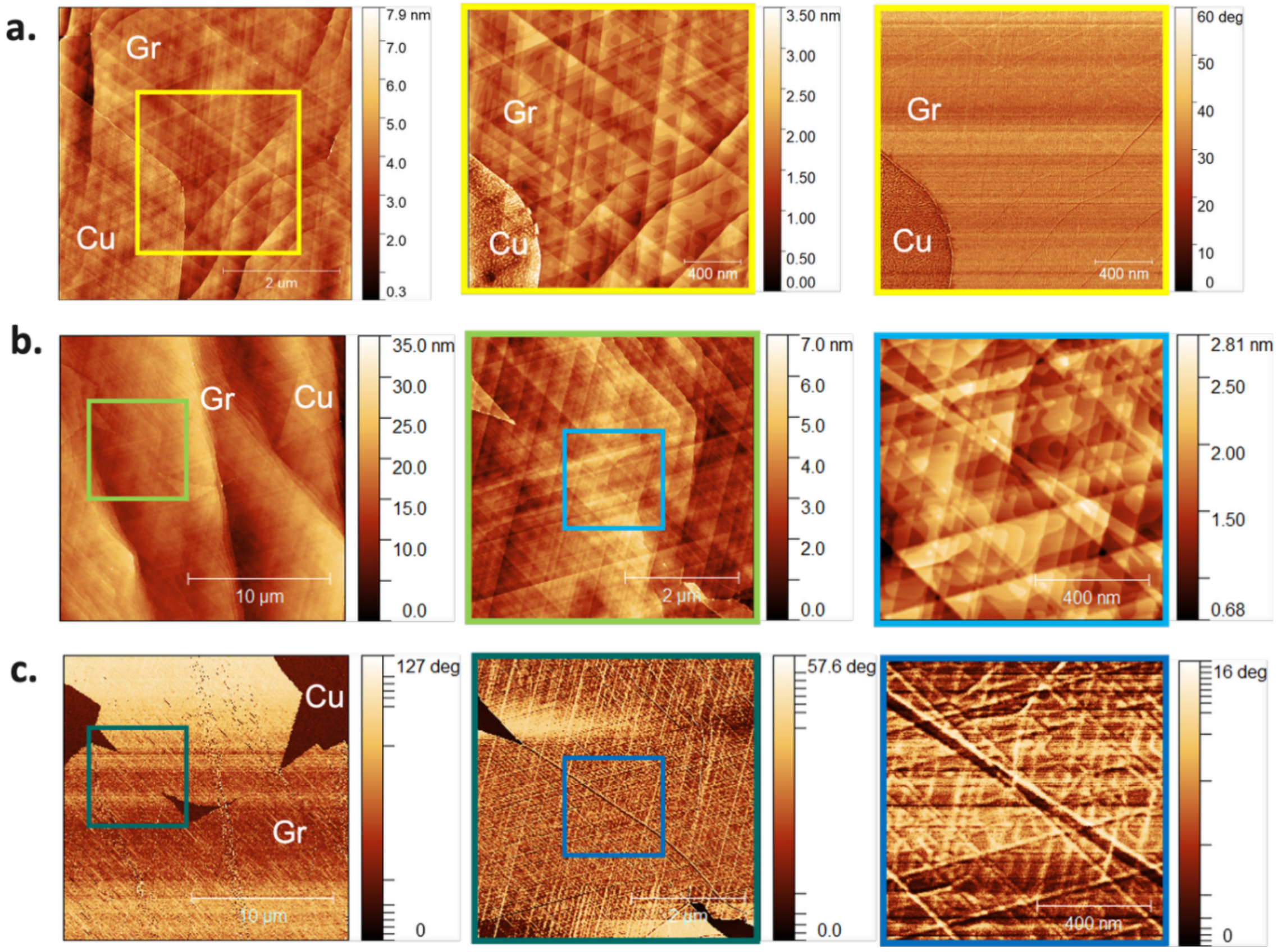

**Figure S3.** Areas with & without the merging lines under AFM. (a) Area 1: merging graphene flakes with no line, (b) & (c) Area 2: topography and phase profile of merging graphene flakes with the merging line

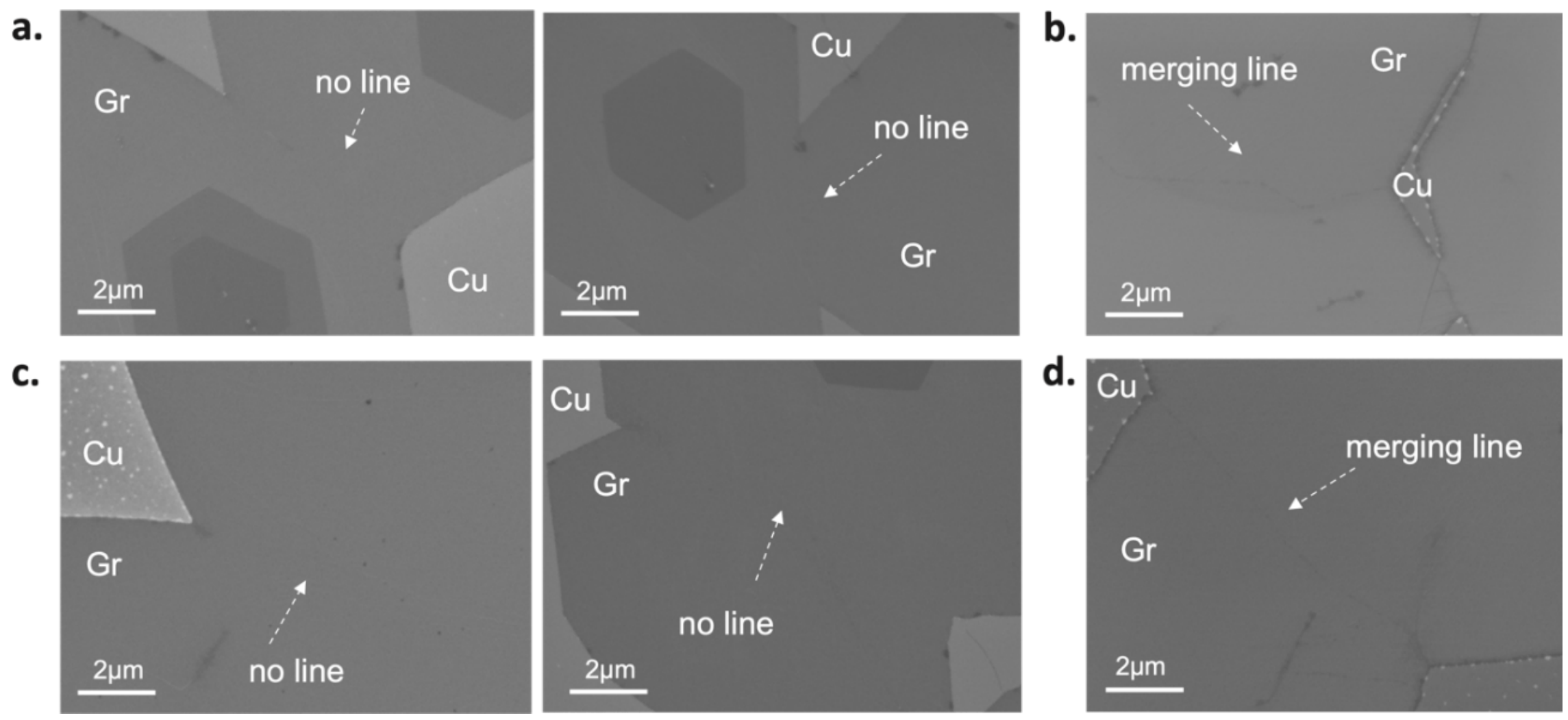

**Figure S4.** Six areas with & without a merging line at the same magnification under SEM. (a)&(c) Four merging areas show no merging line under SEM. (b)&(d) Merging line observed for two graphene flakes stitching together.

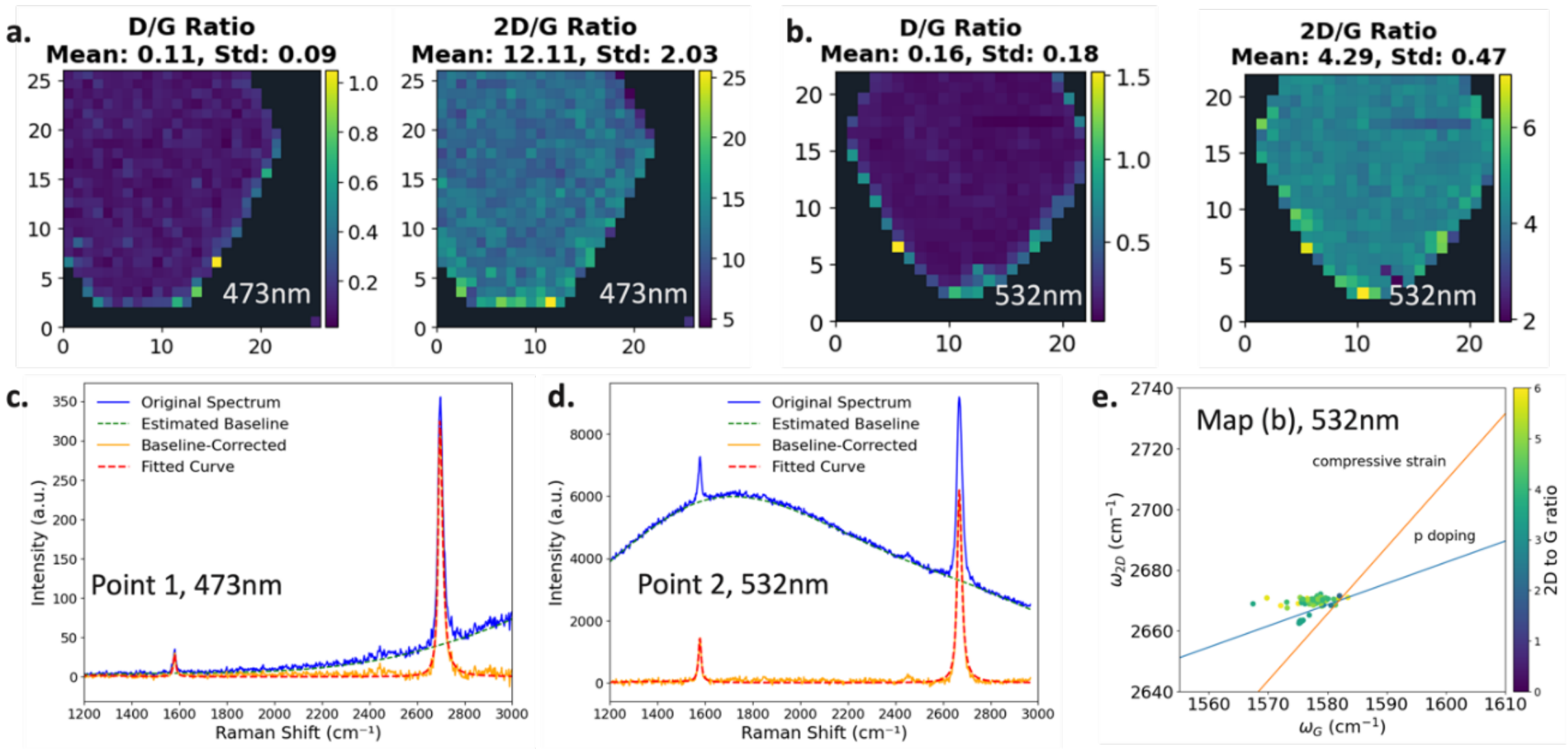

**Figure S5**. Raman mapping of two graphene flakes using 532nm laser and 473nm laser. (a) Mapping of D/G and 2D/G peaks intensity ratio by 473nm laser. (b) Mapping of D/G and 2D/G peaks intensity ratio by 533nm laser. (c) A Raman spectrum of the graphene flake in (a) by 473nm laser, with fittings of G and 2D peaks. (d) A Raman spectrum of the graphene flake in (b) by 532nm laser, with fittings of G and 2D peaks. (e) Raman scattered plot of the mapping data in graphene flake (b). Most points are close to the intersection between the compressive strain line and p doping line, which indicates no compressive strain or p doping. This shows the graphene is decoupled from the copper substrate after wet oxidation.

**Raman maps of position and intensity of G, 2D, and $Cu_2O$ peaks in position 1-3 after wet oxidation**

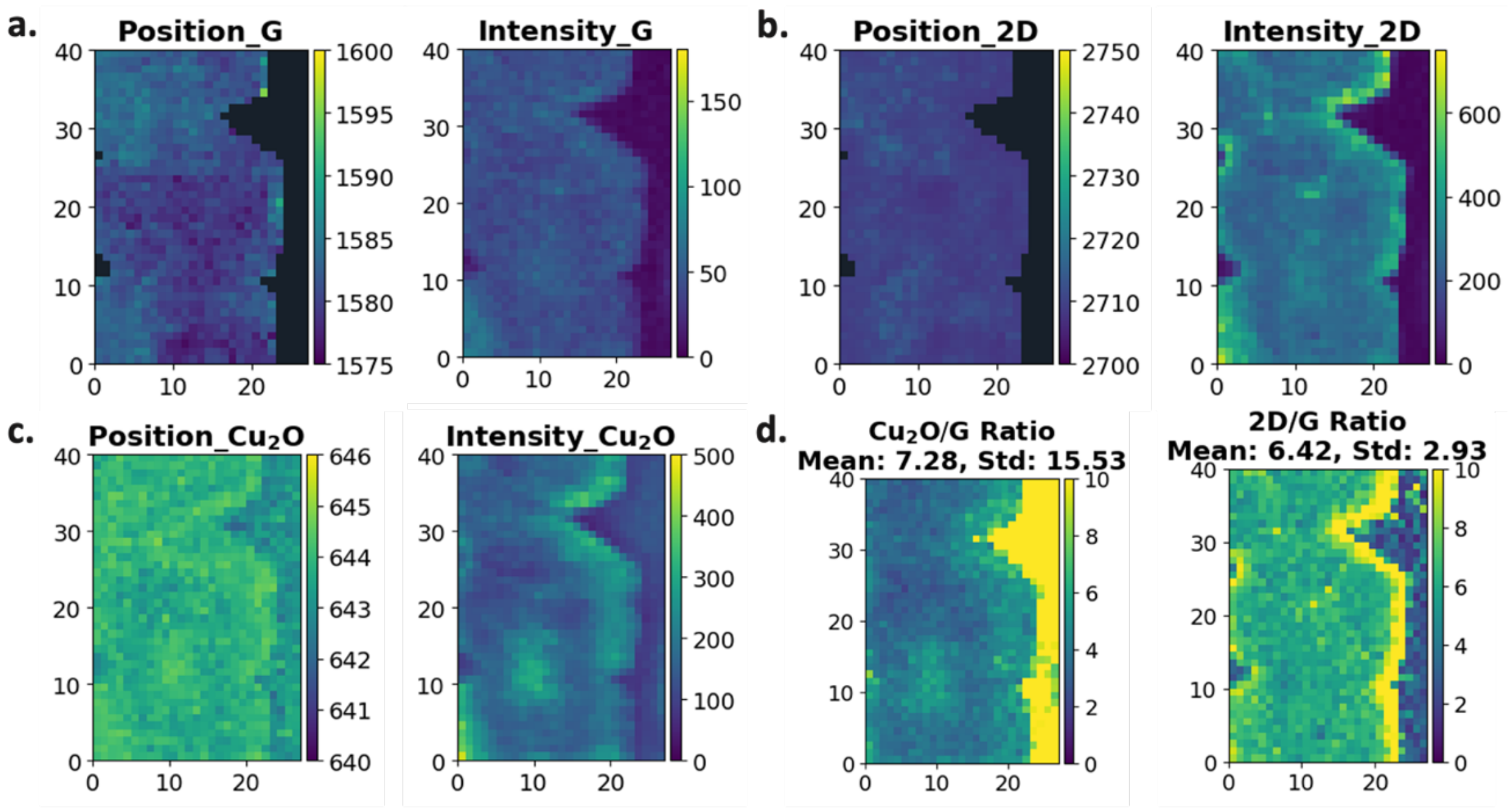

**Figure S6.** Raman mapping of Position 1. Position and intensity maps of (a) G peak. (b) 2D peak (c) $Cu_2O$ peak (d) $Cu_2O$/G and 2D/G peak intensity ratio.

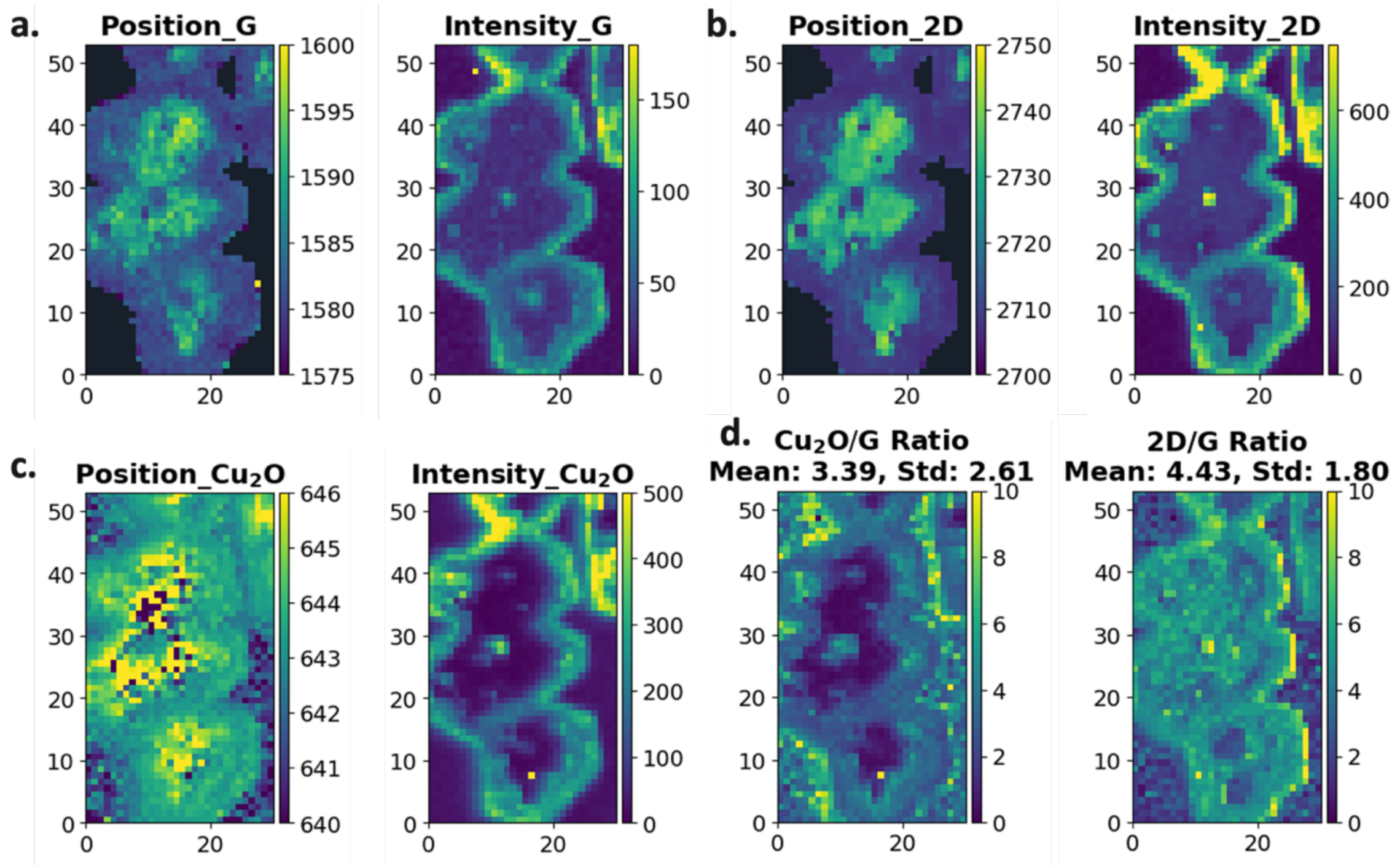

**Figure S7.** Raman mapping of Position 2. Position and intensity maps of (a) G peak. (b) 2D peak (c) $Cu_2O$ peak (d) $Cu_2O$/G and 2D/G peak intensity ratio. For this sample, the center region of the graphene flakes is not oxidized yet, accounting for lack of $Cu_2O$ signal (refer to Figure 2 in the main text for detailed explanation).

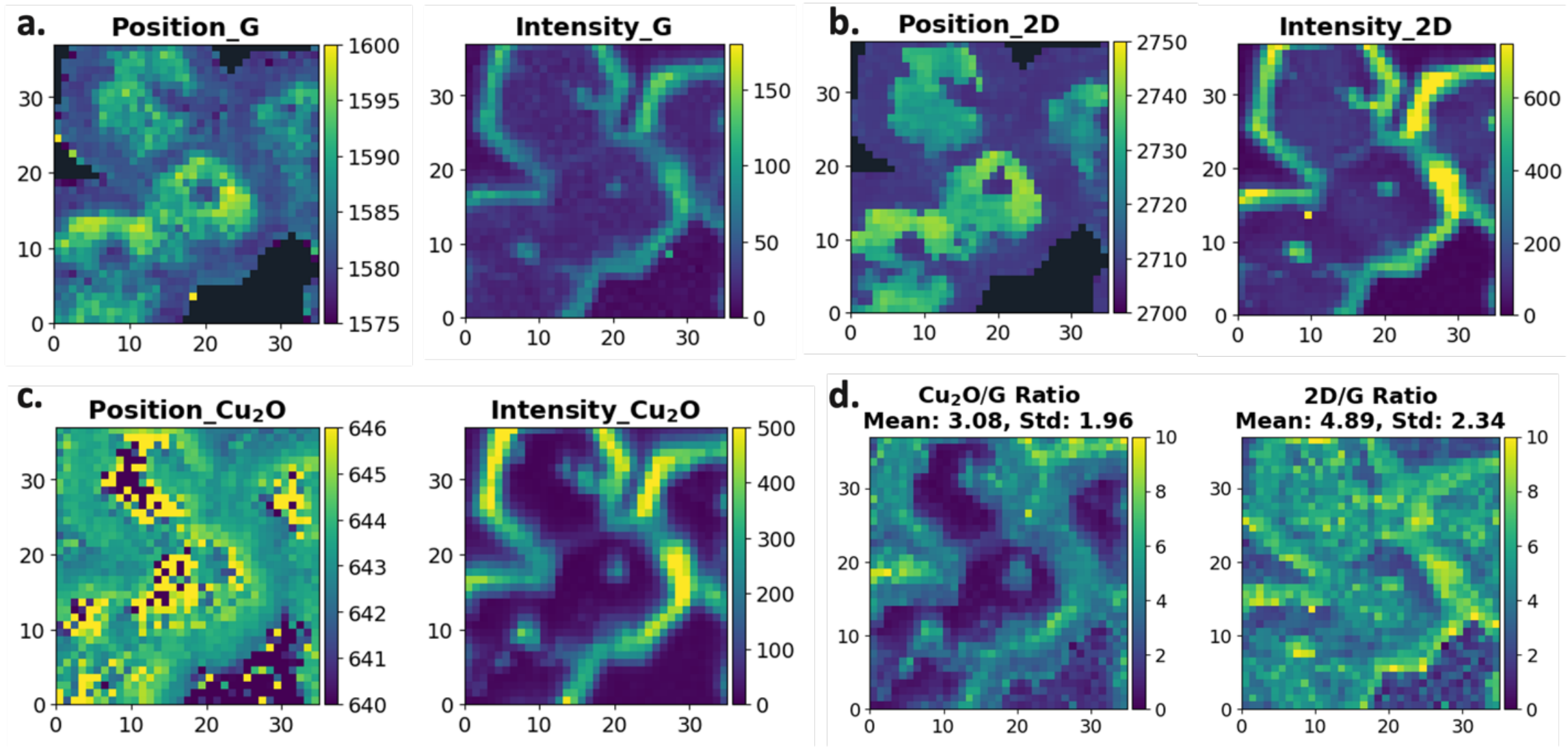

**Figure S8.** Raman mapping of Position 3. Position and intensity maps of (a) G peak. (b) 2D peak (c) Cu2O peak (d) Cu2O/G and 2D/G peak intensity ratio. The center region of graphene flakes is not oxidized in this process, therefore no Cu2O Raman signal (refer to Figure 2 in the main text for detailed explanation).

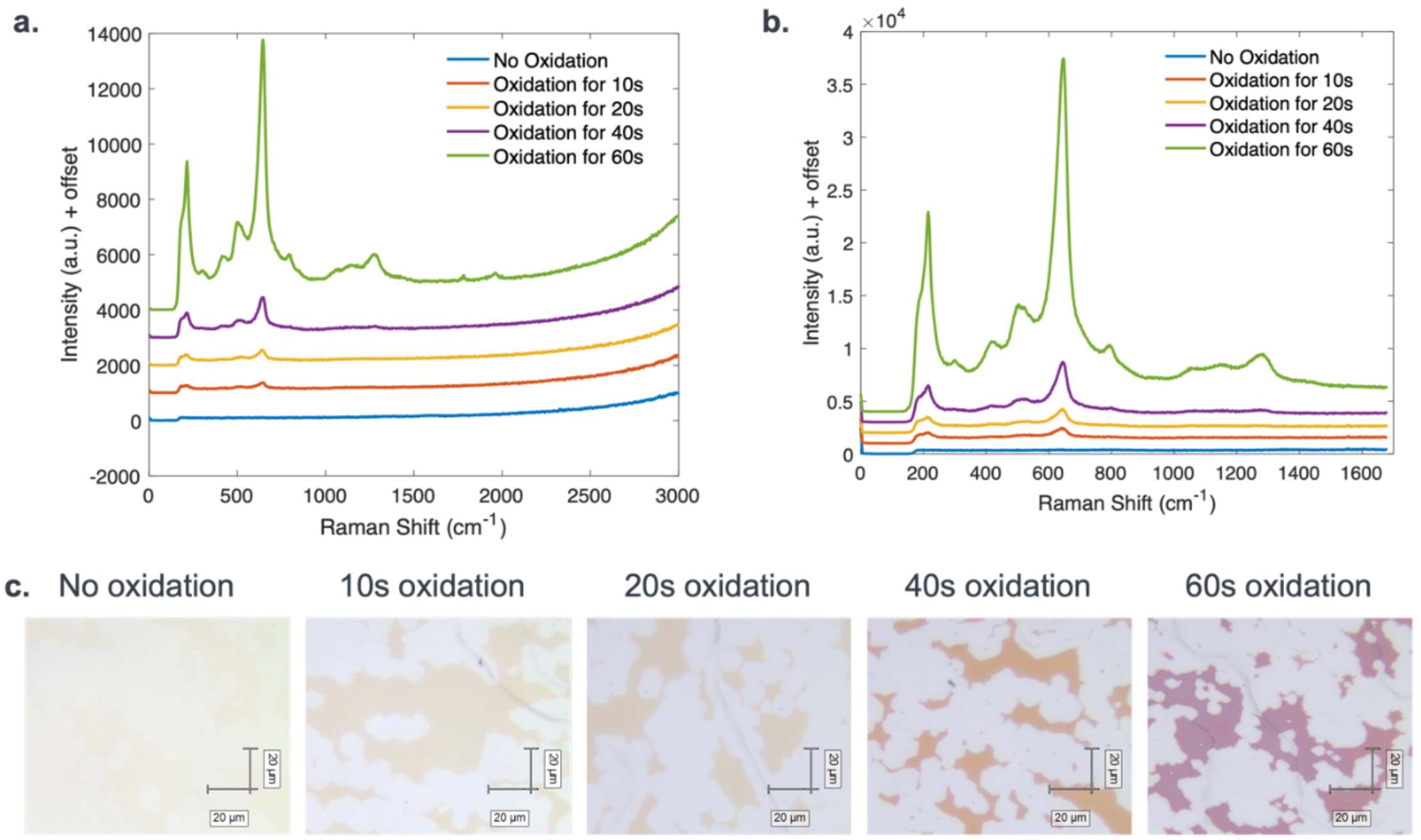

**Figure S9.** Raman spectrum of the Cu(111) film under different oxidation time (a) Raman spectrum of Cu(111) film (not on graphene) under different oxidation time (473nm laser, 1200 grating). (b) Raman spectrum of Cu(111) under different oxidation time (473nm laser, 2400 grating). (c) Optical microscopic image of graphene flakes on Cu(111) film under different oxidation time

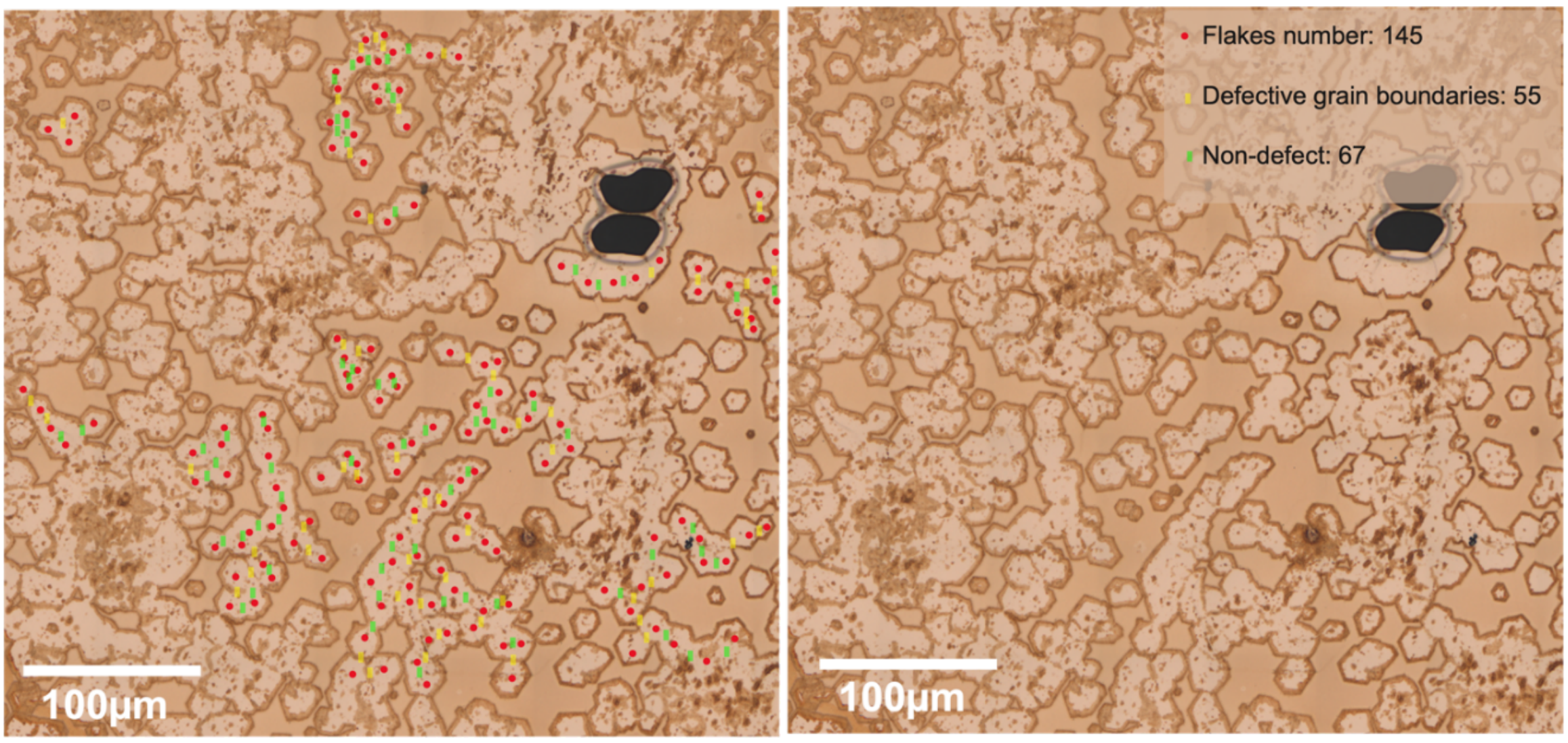

**Figure S10.** Estimation based on an optical microscopic image of defective stitches vs. seamless stitches from 145 merging graphene flakes, after wet oxidation. Left: Optical microscopic image with annotations. Right: original optical microscopic image.

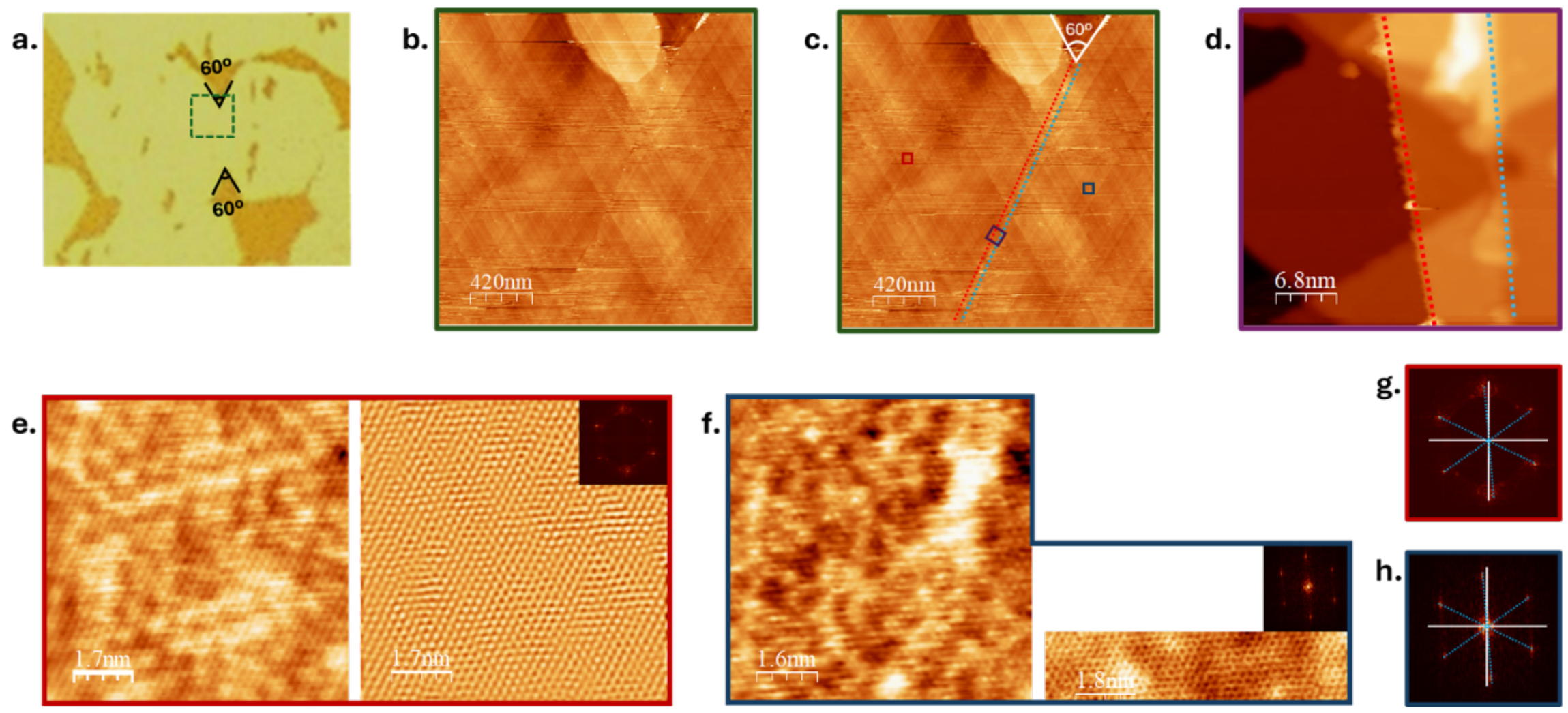

**Figure S11**. STM imaging of the merging line from two aligned graphene flakes.

(a) Optical microscopy image of two merging graphene flakes. The angle between the edges of the graphene flakes, both in optical microscope (a) and STM images (c), is 60°, indicating that both merging flakes are of aligned orientation. (b) STM image of the region outlined by a green square in (a). (c) Same image as (b), with red and blue dashed lines highlighting the overlapped area. The purple square indicates the position of Figure4b. (d) STM image of the purple region at (c), note that it is rotated. The overlapping region (where the two graphene flakes meet) is confined between the dashed lines shown in the image. The red dashed line marks the edge of the upper graphene layer, corresponding to the flake on the right-hand side of the merging line. The blue dashed line indicates the edge of the bottom graphene layer, belonging to the flake on the left-hand side of the merging line (e) Atomic-resolution image on the left graphene flake at the position outlined with a red square in (c). Left, raw data. Right, FFT low pass filtered image highlighting the atomic structure of graphene, the inset being the extracted FFT. (f) Atomic-resolution images on the right graphene flake at the position outlined with a blue square in (c). Inset shows the extracted FFT. (g),(h) Comparison between the extracted FFTs for atomic-resolution images at both sides of the overlapped region. They show the same atomic orientation, further confirming that the merging flakes are aligned.

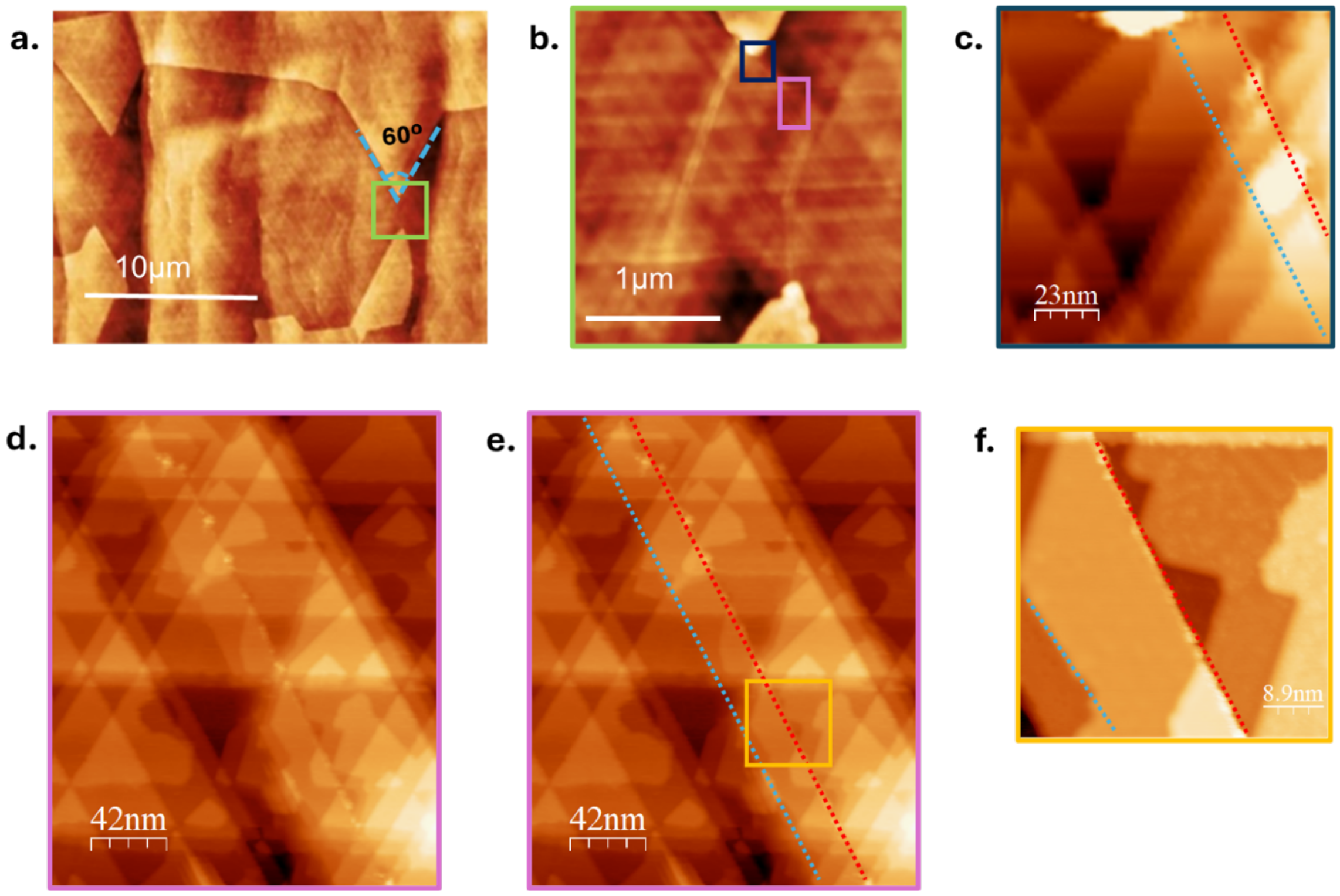

**Figure S12.** STM imaging of merging lines. (a) AFM image of two merging graphene flakes with aligned orientation, as indicated by the 60° angle between the edges of both graphene grains, outlined at the top of the merging region. (b) AFM image of the region outlined by the green square in (a). The two rectangles, pink and blue, indicate the two areas of the merging lines characterized by STM, Figure 4e-g and Figure 4h-k respectively. (c) STM image of the overlapping area at the blue square in panel (b), corresponding to Fig 4h. The overlapping region is confined between the dashed lines shown in the image. The red dashed line marks the edge of the upper graphene layer, corresponding to the flake on the left-hand side of the merging line. The blue dashed line indicates the edge of the bottom graphene layer, belonging to the flake on the right-hand side of the merging line. (d) STM image of the region outlined by a pink rectangle in (a). (e) Same image as (d), with red and blue dashed lines highlighting the overlapped area. The yellow square indicates the position of Figure4e. (f) STM image at the yellow square position in (e). The overlapping region is confined between the red and blue dashed lines outlining its borders.

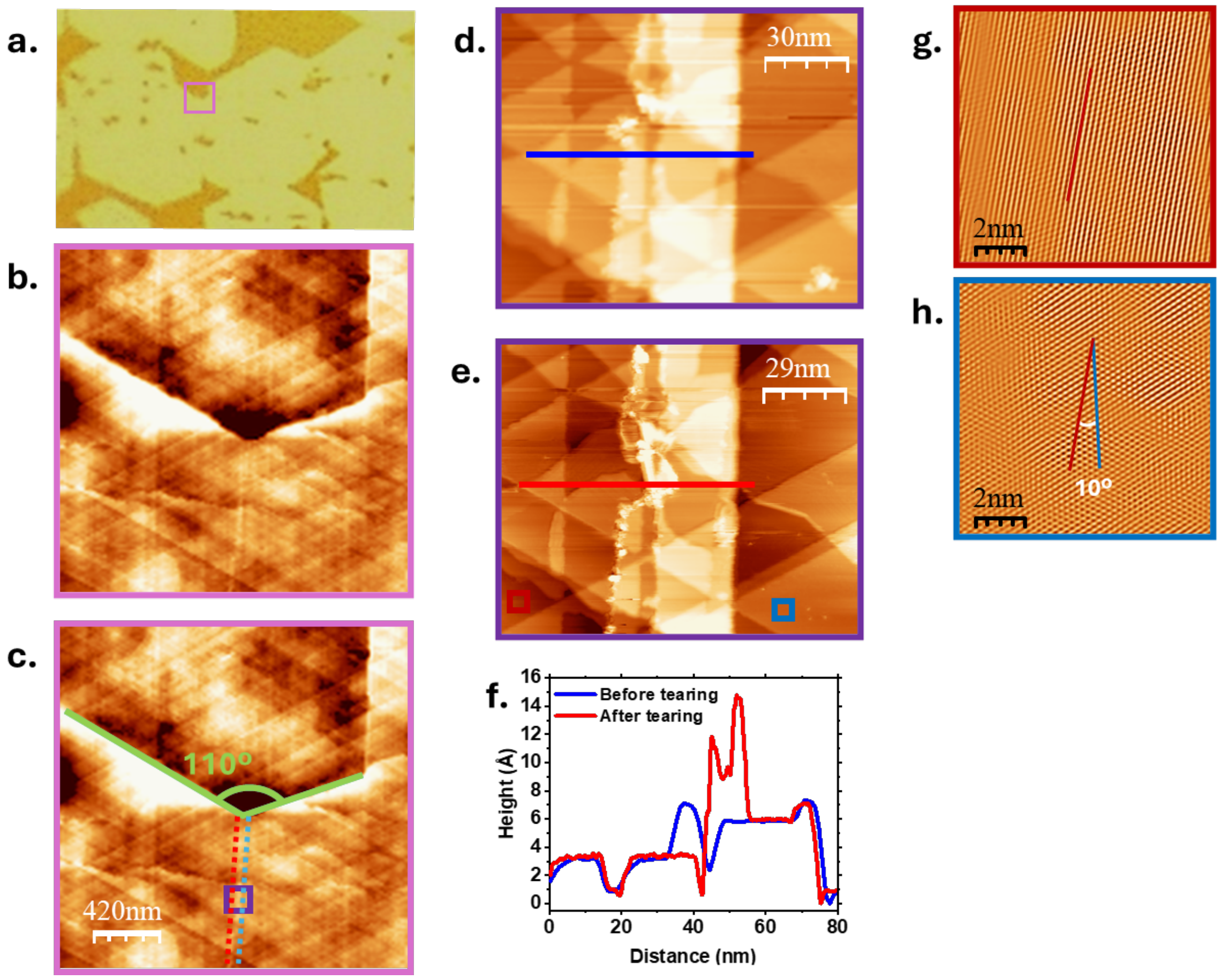

**Figure S13**. STM imaging of the merging line of two graphene flakes with 10° mismatch in orientation. (a) Optical microscope image of two misaligned merging graphene flakes. (b) STM image of the region outlined by a pink square in (a). (c) Same image as (b). The angle between both merging graphene flakes form 110º, indicating a 10-degree misalignment. Red and blue dashed lines highlight the overlapping region (d) STM imaging of the area indicated as a purple box in (b), an overlapping region of around 30-50nm wide is clearly seen. (e) Tearing happens during STM scanning over a highly defective region of the upper edge of the merging line. (f) Height profile across the merging line measured before (d) and after (e) the tearing. The step height measurements across the merging line support the structure of an overlapped junction. (g), (h) FFT Filtered STM images highlighting graphene's atomic structure from both merging flakes, measured at the positions indicated in (e) by the red and blue squares. The 10° misalignment between the atomic lattices of both graphene flakes matches the 10° misalignment angle observed between their edges in (c).

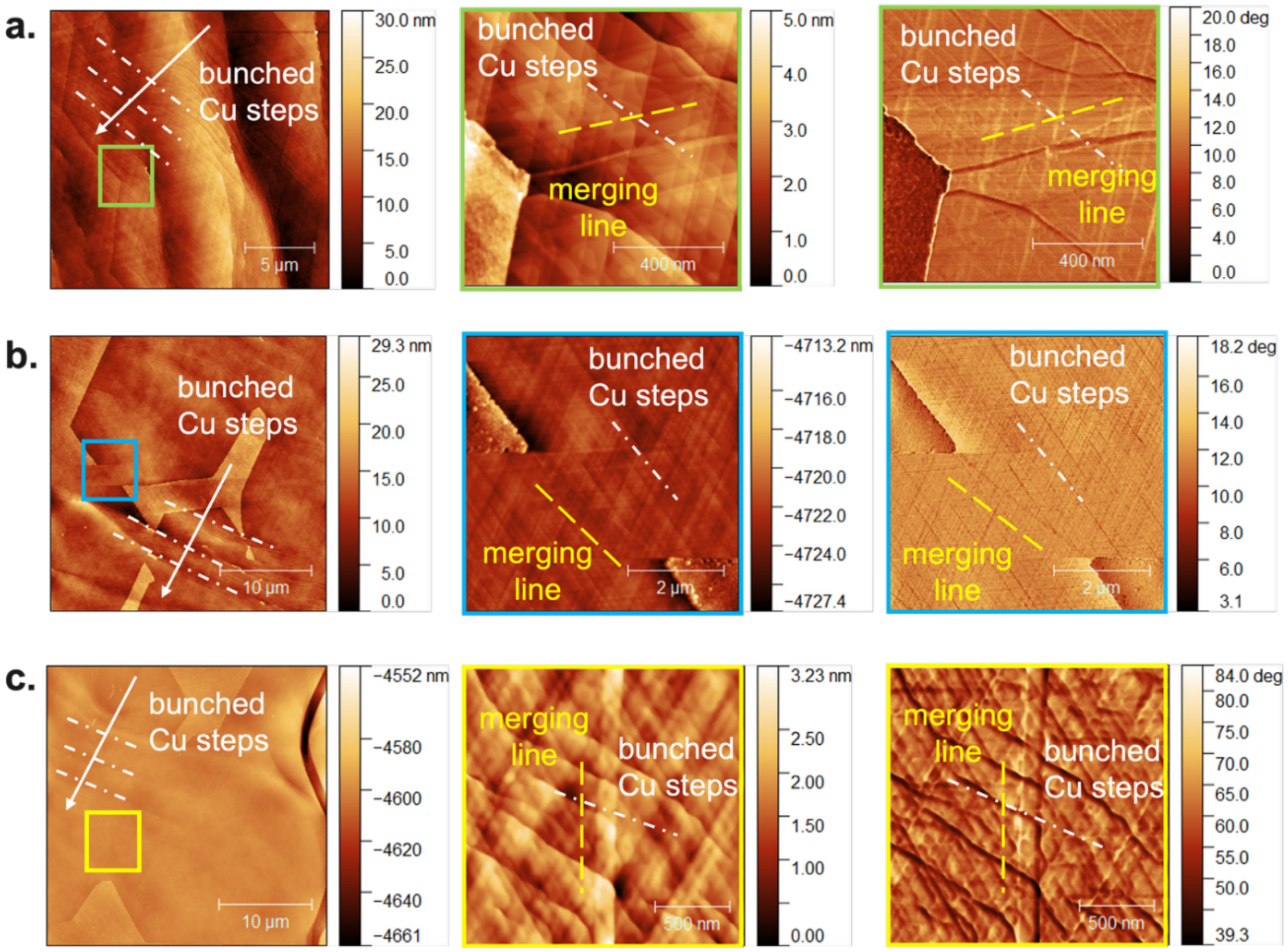

**Figure S14.** Merging lines in different angles with regards to the direction of bunched Cu steps. (a)&(c) AFM image of two merging graphene flakes, where the merging line is in a small angle with the direction of bunched Cu steps. (b) AFM image of two merging graphene flakes, where the merging line is in nearly 90 degrees with the direction of bunched Cu steps.

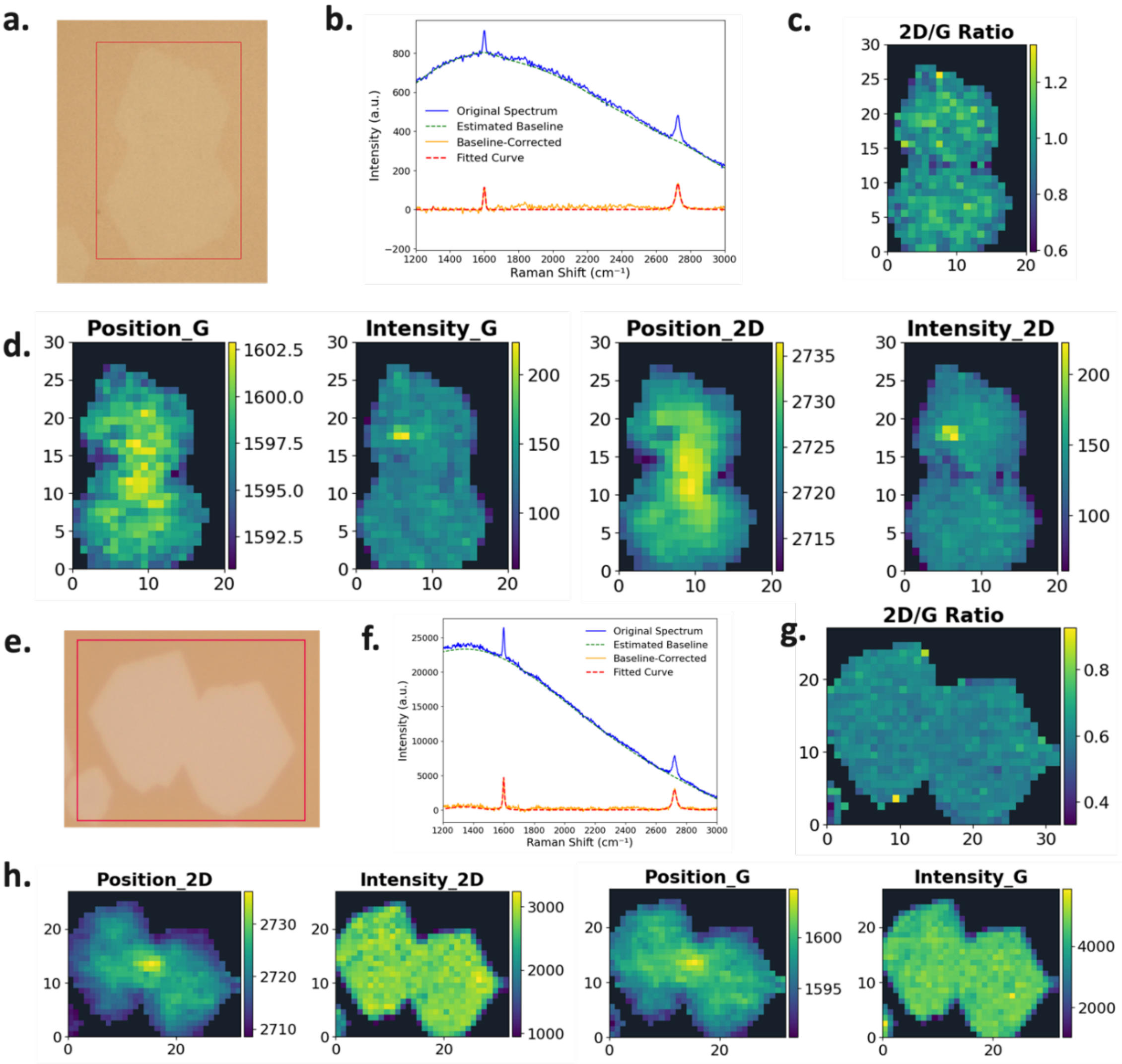

**Figure S15.** Higher compressive strain at the merging area of two aligned graphene flakes. (a) Area 1 of two merging graphene flakes. (b) Raman spectrum of one spot in area 1 with fitting. (c) Raman map of 2D/G peak intensity ratio on area 1. (d) Raman map of G and 2D peak position and intensity of area 1. (e) Area 2 of two merging graphene flakes. (f) Raman spectrum of one spot in area 2 with fitting. (g) Raman map of 2D/G peak intensity ratio on area 2. (h) Raman map of G and 2D peak position and intensity of area 2. Red-shift of both G and 2D peak positions at the merging area, indicating higher compressive strain at the merging area. Raman map in area 1 is obtained with a lower laser power to preserve the graphene for further characterization, which accounts for the lower intensity in G and 2D peaks.